**Can ethnic tolerance curb self-reinforcing school segregation? A theoretical Agent Based Model**

**Authors:**

**Lucas Sage - Sorbonne Université, GEMASS; University of Trento**

**Andreas Flache - University of Groningen, ICS**

**Abstract**

Schelling and Sakoda prominently proposed computational models suggesting that strong ethnic residential segregation can be the unintended outcome of a self-reinforcing dynamic driven by choices of individuals with rather tolerant ethnic preferences. There are only few attempts to apply this view to school choice, another important arena in which ethnic segregation occurs. In the current paper, we explore with an agent-based theoretical model similar to those proposed for residential segregation, how ethnic tolerance among parents can affect the level of school segregation. More specifically, we ask whether and under which conditions school segregation could be reduced if more parents hold tolerant ethnic preferences. We move beyond earlier models of school segregation in three ways. First, we model individual school choices using a random utility discrete choice approach. Second, we vary the pattern of ethnic segregation in the residential context of school choices systematically, comparing residential maps in which segregation is unrelated to parents' level of tolerance to residential maps reflecting their ethnic preferences. Third, we introduce heterogeneity in tolerance levels among parents belonging to the same group. Our simulation experiments suggest that ethnic school segregation can be a very robust phenomenon, occurring even when about half of the population prefers mixed to segregated schools. However, we also identify a "sweet spot" in the parameter space in which a larger proportion of tolerant parents makes the biggest difference. This is the case when the preference for nearby schools weighs heavily in parents utility function and the residential map is only moderately segregated. Further experiments are presented that unravel the underlying mechanisms.

## 1.    Introduction

School segregation is a threat to integration in multicultural societies. The clustering of different ethnicities in different schools is a persistent phenomenon in many countries (Bakker 2011). It contributes to the widening of educational, occupational, and earnings inequalities between ethnic groups (Ashenfelter, Collins, and Yoon 2006; Johnson 2011; Reardon and Owens 2014). Furthermore, schools have a key role to play in forming young generations' attitudes toward diversity. Indeed, in the wake of (Allport 1954) a large stream of research has demonstrated that under favorable conditions, interethnic group contacts can powerfully reduce ethnic prejudices (Christ et al. 2014; Powers and Ellison 1995). Intergroup contacts are even more important during youth, when children's and adolescent's attitudes are still forming (Wölfer et al. 2016). Therefore, early positive interethnic contacts are crucial to form the attitudes next generations of citizens' have towards diversity. However, school segregation strongly limits opportunities for interethnic context in the first place.

While school segregation can have highly undesirable societal consequences, there still is much to be learned about the social mechanisms causing it. Interestingly, a recent sociological literature review on school segregation (Reardon and Owens 2014) shows that the overwhelming majority of studies have assessed its consequences for educational inequality rather than its causes. Some studies, however, do address the issue by either documenting micro-level parental preferences for school characteristics through surveys or interviews (Borghans, Golsteyn, and Zölitz 2015; Burgess et al. 2007; Karsten et al. 2003), or by looking at the relationships between school and residential segregations at the macro level (Frankenberg2013; Johnston et al. 2006). Yet, what is largely absent in this kind of work is the connection between micro-level parental choice behavior and macro-level school segregation patterns. One reason why such work is much needed in our view is that, as a recent review of the field highlights, school segregation outcomes can strongly differ between different spatial and institutional contexts (Boterman et al, 2019). Our paper contributes to a small but growing literature addressing by means of agent-based modelling the question how interdependent parental school choices aggregate to produce school segregation in different social and spatial contexts, and under which conditions school segregation could be curbed.

A better understanding of the mechanisms driving segregation is particularly relevant because of the large-scale increase of ethnic diversity many societies are currently experiencing. Some even see a "diversity explosion" among the millennial generation (W. Frey 2018; W. H. Frey 2018; Tasan-Kok et al. 2014; Vertovec 2007), raising hopes of more ethnic integration and of the erosion of ethnic boundaries in the foreseeable future. Notably, these new generations are populated by a growing number of mixed-race households (Ellis et al. 2007, 2012), and mixed-race individuals (Clark and Maas 2009), which have stronger preference for multicultural neighborhoods (Clark, Andersson, and Malmberg 2018). A very recent study specifically analyzing residential choices of the millennials (Clark and Brazil 2019) shows considerable individual variation among ethnical groups, but also lower propensity to remain in ethnically homogenous neighborhoods dominated by co-ethnics. In Europe surveys show that large fractions of natives are relatively tolerant towards ethnic minorities (Andersson, Brattbakk, and Vaattovaara 2017).

The rise of diversity is accompanied by considerable transformation in residential segregation patterns. In the USA, (Glaeser and Vigdor 2012) announced "the end of the segregated century", and highlighted that entirely white neighborhoods almost completely disappeared. Others (Alba and Romalewski 2013) have a more nuanced view but many analysts converge to the idea that neighborhoods are becoming increasingly ethnically diverse (Farrell and Lee 2011; Hall, Tach, and Lee 2016; Lee, Iceland, and Sharp 2012). The emergence of ethnically mixed neighborhood is according to (Clark 2015) the major change in residential ethnic segregation of the past decades.

The rise of ethnic tolerance among significant fractions of the population can be expected to also have profound impact on school segregation. However, formal modelling work raises doubts as to whether and, if so, to what degree more tolerance among parents can really change the strong self-reinforcing tendencies of segregation which Schelling's and Sakoda's (Hegselmann 2017; Sakoda 1971; Schelling 1971) famous models prominently uncovered. Schelling's model in particular is probably one of the best-known examples of a process of social self-organization in which multiple interdependent individual actions generate an aggregate outcome which individual actors neither intend nor desire to bring about. These models famously demonstrated how neighborhoods can end up strongly ethnically segregated due to "preference dynamics" (Clark and Fossett 2008), even when individuals have largely tolerant ethnic preferences. A large follow-up literature has shown this result to be robust across a wide range of variations of the model and contextual conditions under which it is applied (Clark and Fossett 2008; Fossett 2006; van de Rijt, Siegel, and Macy 2009). Recent work (Stoica and Flache 2014) demonstrated how the argument of preference dynamics also extends to the emergence of unintended school segregation. Yet, this work gives little insight into the role of increasing variation in ethnic tolerance as we currently witness it in modern societies.

Could higher levels of ethnic tolerance change the seemingly inevitable process through which segregated neighborhoods and schools arise under preference dynamics? Schelling himself pointed to the opposite possibility. Heterogeneity in preferences may be a crucial factor to destabilize ethnically mixed neighborhoods. This idea was elaborated in the "bounded neighborhood model" or "tipping model" (Schelling 1971), in which cascades that lead to the unravelling of integrated neighborhoods are triggered by a few highly intolerant agents who are dissatisfied with the local ethnic mix and therefore move out first, followed by slightly less intolerant individuals, and so on (Skvoretz 2006). Extending this argument to schools, one might expect that heterogeneity in tolerance will destabilize ethnically diverse schools, because of the behavior of intolerant parents.

On the other hand, recent work that explicitly included heterogeneous ethnic preferences in agent-based models of residential Schelling-type dynamics, suggests that the implications of heterogeneous preferences may be less straightforward (Fossett 2006; Hatna and Benenson 2015; Xie and Zhou 2012). Xie and Zhou (2012) extended the Schelling-type model with heterogeneous preference distributions. They found that segregation levels are in the long run even lower than they would be under homogeneous preferences. The reason is that more tolerant agents self-sort into relatively mixed neighborhoods, stabilizing the local integration and filling the vacancies left behind by intolerant agents who moved out. Extending this argument to schools one could expect heterogeneous patterns, with some schools integrated and some others segregated, in the presence of large proportion of tolerant parents in an ethnically diverse population.

Thus, the literature points to conflicting intuitions as to the consequences of tolerance heterogeneity among parents for school segregation. Deciding which intuition is more plausible becomes even more complicated because preference dynamics in school choice are not just a mirror image of those postulated by models of residential segregation. Building upon empirical work on school choice preferences (Borghans et al. 2015; Burgess, Wilson, and Lupton 2005; Karsten et al. 2003), Stoica and Flache (2014) integrated a new element relevant for school segregation into a model of preference dynamics, individuals' preferences for geographical proximity of schools. They assumed that - all other things being equal - parents favor nearby schools over distant schools. This could curb school segregation by keeping intolerant parents from abandoning ethnically diverse schools in their neighborhood. But distance preferences could also foster segregation by keeping tolerant parents living in segregated neighborhood from sending their children to mixed schools further away from home. One might expect complex interplays between heterogeneity in parents' ethnic tolerance, residential segregation patterns and the strength of parents' preference for nearby schools.

To clarify which intuition might turn out to be true, and under which conditions, we develop a new model of school choices that moves beyond the previous work in three important ways. First, following recent work on residential segregation dynamics (Hatna and Benenson 2015; Xie and Zhou 2012), we take into account that there is heterogeneity among parents in the degree to which they are ethnically tolerant. Simulations reveal that the introduction of a substantial proportion of tolerant parents in a population can crucially affect which pattern of school segregation eventually emerges. Second, we consider variation in the residential structures that form the context of school choice. Finally, we drop the assumption inherited from Schelling and Sakoda and used in most modelling work, that actors' satisfaction with a school or residential location is fully determined by its characteristics. Instead, our new model links up with recent advances in models of residential segregation (Bruch and Mare 2006; van de Rijt et al. 2009) and uses a random utility specification of choice behavior where the probability of selecting a school is positively associated with the potential satisfaction of an agent (McFadden 1973; for an empirical modelling application see: Borghans et al. 2015), while random deviations due to unobserved heterogeneity are also considered.

The paper is organized as follows: section 2 details the three principal modeling assumptions we make; section 3 describes our formal model; section 4 shows results of our simulation experiments; section 5 discusses implications of our results and possible avenues for future research.

## 2. Towards a more realistic model of school segregation dynamics: heterogeneity in tolerance, mixed residential distributions and random choice behavior

In this section, we discuss the possible impacts of heterogeneity in parents' tolerance levels (section 2.1), variation in residential segregation patterns (section 2.2) and non-deterministic choice preferences (section 2.3) on school segregation dynamics. We then develop a modeling strategy to analyze how a larger proportion of tolerant parents in the population interacts with residential segregation patterns and parents' preferences for nearby schools on the level of school segregation our model generates.

### 2.1. Heterogeneity of ethnic preferences: empirical evidences and theoretical consequences

The integration of heterogeneous ethnic preferences into our model is motivated by empirical research showing considerable heterogeneity in school choice preferences not only between but also within different ethnic and social groups (Borghans et al. 2015; Hastings, Kane, and Staiger 2006). While these studies measure heterogeneity not explicitly for ethnic preferences but focus instead on distance preferences, preferences for high school quality, or for a particular school denomination, heterogeneity in ethnic preferences has for example been demonstrated by factorial survey studies of parents' resistance against schools with varying proportions of outgroup members (Billingham and Hunt 2016; Coenders, Lubbers, and Scheepers 2004; Goyette, Farrie, and Freely 2012).

A further reason why ethnic preferences should be assumed to be heterogeneous is that parents' preferences for school ethnic composition are likely to mirror preferences for neighborhood ethnic composition. Using increasingly sophisticated measurement instruments, studies of residential ethnic preferences clearly provide evidence of heterogeneity. For example, based on data from the Multi-City Study of Urban Inequality (MCSUI), (Xie and Zhou 2012) find that respondents can be classified into 6 different categories between which the data show significant differences in the preferences they have for the proportion of ethnic outgroups present in an acceptable neighborhood. This also reflects results from a number of other studies on residential preferences (Clark and Fossett 2008; Farley et al. 1978; Ibraimovic and Masiero 2014) and residential choices (Clark and Brazil 2019).

As discussed above, recent modelling work of residential segregation dynamics has shown how heterogeneity in tolerance can considerably change the outcomes and the process of ethnic preference dynamics (Hatna and Benenson 2015; Xie and Zhou 2012). Therefore, in the present study, we systematically explore under what conditions heterogeneity in ethnic preferences affects the self-sorting of tolerant parents into integrated schools occurs and how this interacts with preferences for nearby schools.

## 2.2.    Importance of the residential context

Previous modelling work also has left unexplored how the residential context of school choice has an impact on emergent ethnic school segregation. For example, whether a preference for nearby schools fosters ethnic school integration and curbs self-reinforcing school segregation, may critically depend on residential segregation. Preferences for nearby schools would reduce school segregation in a residentially integrated context, as schools' ethnic compositions tend to reflect the composition of neighborhoods if most parents prefer nearby schools (Stoica and Flache, 2014). Yet, real world cities are far from being perfectly integrated (Farley and Taeuber 1968, 1974; Frankenberg 2013; Johnston et al. 2006; Ong and Rickles 2004; Reardon and Yun 2005).

The consequences of the degree of segregation in the residential setting could however be less straightforward at the local level. Consider schools that are situated close to boundaries between ethnically homogeneous residential clusters. Their catchment area would contain all the closest dwellings, i.e. parents from both sides of the ethnic frontier. These schools would therefore initially be highly integrated. This points to the possibility that strong distance preferences may also foster school integration in some specific schools in a residentially segregated world, because ethnically intolerant parents could be reluctant to bear the cost of sending their children to more segregated but more distant schools. The interaction of distance preferences and heterogeneity in tolerance might also explain the emergence of a mix of segregated and integrated schools similar to the mixed residential patterns observed in many real-world cities. In short, while intolerant parents would try to get away from integrated schools and be willing to choose more distant segregated schools, tolerant parents may be willing to bridge larger distances to find ethnically integrated schools. This suggests a potential for a complex interplay between two interdependent yet different forms of segregation. Schools can be segregated by ethnicity, but they can also be segregated by tolerance. Ethnically segregated schools would be populated by more intolerant parents, while more tolerant parents would choose ethnically integrated schools. Our simulation experiments will explore how this possibility of simultaneous segregation by tolerance and by ethnicity interacts with the residential map in a city.

## 2.3. Probabilistic school choice

Following Schelling's and Sakoda's classic formulations, most formalizations of Schelling-type segregation dynamics assume that individuals will never abandon a residential neighborhood or school as long as its ethnic composition remains satisfactory in the light of their preferences (Hatna and Benenson 2015; Stoica and Flache 2014). As a consequence, such models can generate "frozen states", such as stable ethnically integrated schools or neighborhoods, which would collapse under more realistic assumptions allowing occasional random moves or "erroneous" individual choices. In order not to overestimate the potential of ethnic tolerance or distance preferences for stabilizing ethnically integrated distribution, we adopt in line with recent modeling work on residential segregation (Bruch and Mare 2006; van de Rijt et al. 2009; Zhang 2004) a random utility model. In such a model, the satisfaction given by a school is positively associated with the probability of selecting it, but parents do not always opt for the most satisfactory option. Substantively, this randomness can be interpreted as resulting from heterogeneity in unobserved factors other than distance and ethnic mix which affect parents' evaluation of a school, but also to some dissonance between preferred and actual school or neighborhood (Schwanen and Mokhtarian 2004).

## 3. The model

We propose a model that uses largely the same Schelling-Sakoda type framework for generating both the residential segregation pattern and the pattern of school choices parents make. Section 3.1 explains the resulting two-steps procedure, section 3.2. elaborates the decision-making model, 3.3 describes how residential maps are generated and where schools are located, and section 3.4 formalizes the outcome measures quantifying emergent segregation.

### 3.1. General logic of the simulations: a two steps procedure.

We proceed in our simulations in two steps. In the first step, a residential map with varying degree of segregation is generated using a Schelling-type algorithm. At the end of this step, parents stop moving for the rest of the simulation. In the second step, we first generate schools at random positions (on the residential map generated in the first step), and then assign parents to the closest school from their fixed residential location, corresponding to a Voronoi diagram (Flache and Hegselmann 2001; Okabe, Boots, and Sugihara 1992). This creates schools with initial ethnic compositions that are representative of their neighborhood. From this initial situation, the school choice process starts: parents evaluate the attractiveness of available schools based both on their distance, and on their ethnic composition.

Our model examines "post-residential" school choice (Hastings et al. 2006), neglecting the complication that residential choice could also be influenced by school choice. Given there are hardly any models in the literature so far that bridge the dynamics of residential segregation and school segregation, we believe this simplification is necessary. We also believe it is warranted as a first approximation, because residential choice is subject to many more constraints than school choice, including housing prices, distance to work, the composition of the neighborhood, and the need to satisfy all members of a household simultaneously (Kim, Pagliara, and Preston 2005; Schwanen and Mokhtarian 2004). Thus, we make the simplifying assumption that once parents have chosen a house, they select between available schools rather than changing their dwelling.

In what follows, the school choice model is described first, because it is the main model of interest in our study. The choice model used to generate the residential maps will then be derived as a simplified form of the school choice model, and more details are given in appendix A.

### 3.2. From utility to school choice: decision making process of parents

We use a two-dimensional grid with 6400 cells (80x80) embedded on a torus. 90% of the cells (5760) are populated by one "household" which is the decision-making unit modelled by our agents. Households have a fixed residential location. Once agents have been assigned to residential locations, 30 schools are randomly placed on the map. For simplicity, each household is assumed to have only one child. Figure 1 depicts three examples of residential maps and corresponding school assignment graphs. The household's school choice is represented as a tie between a household and a school in a bipartite graph, with households and schools as the two types of nodes (right panel). A school choice decision can either be maintaining the already existing tie, *i.e.* stay, or erase it and replace it by another, *i.e.* change. Every household is linked to exactly one school at any point in time. Initially all schools have free capacity to receive new pupil, but schools also have a limited maximum capacity which is the same for all schools and is equal to 403 pupils (7% of the total population). If a school has reached its maximum capacity it is no longer part of the possible choice set of parents, until one pupil leaves. This maximum capacity parameter is of importance in real-world settings. For example, (Kessel and Olme 2018) show that the rule determining admission in schools that face higher demand than seats can impact segregation levels. Further below, we present some simulation results exploring effects of variation in the assumptions about maximum capacity.

For simplicity, we assume that the population of households consists of two ethnic groups of equal sizes. Diversity in ethnic preferences is introduced by dividing both ethnic groups into two different categories, tolerant and intolerant agents. For tolerant agents, the ideal school comprises a significant share of outgroup members, whereas intolerant agents prefer schools with very few or even no outgroup members, all other things (distance) being equal.

A:   integrated map

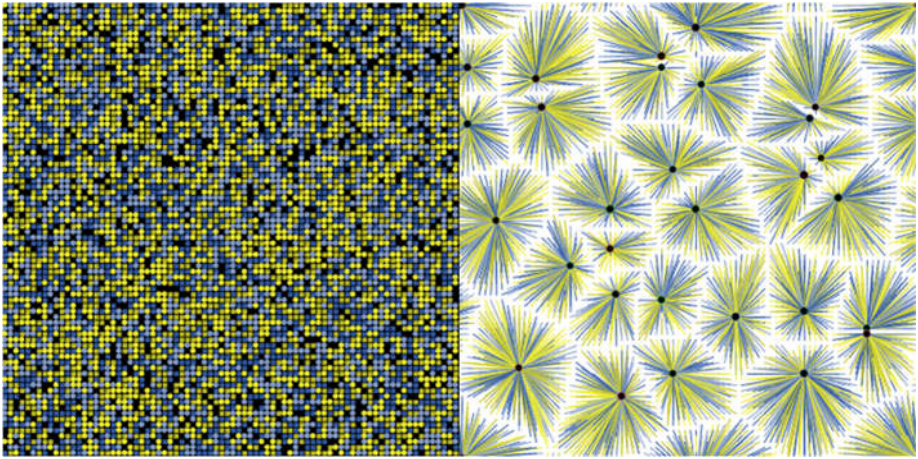

B:     simple
segregation

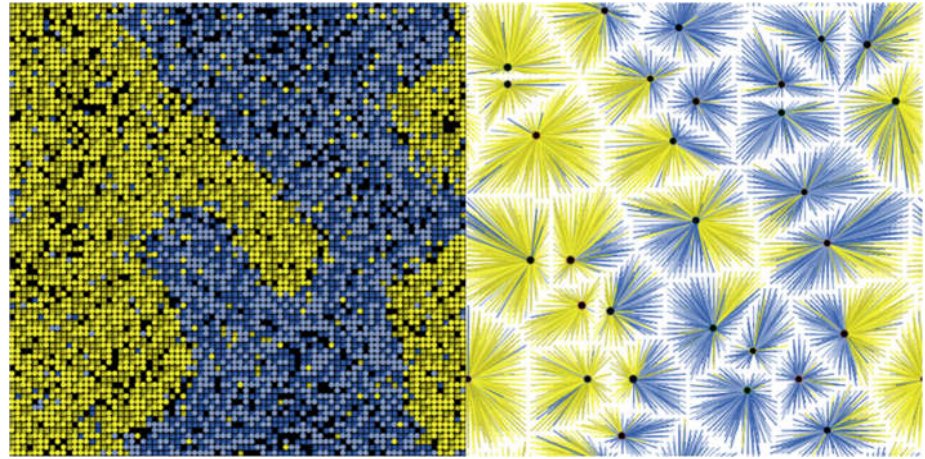

C:     complex
segregation

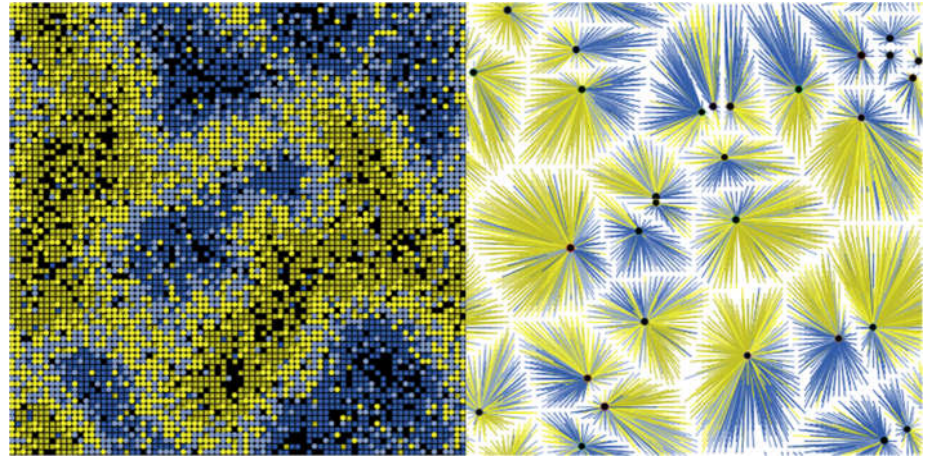

🟨 Intolerant parent, green ethnicity        🟡 Tolerant parent, green ethnicity
🟦 Intolerant parent, blue ethnicity         🔵 Tolerant parent, blue ethnicity
⬛ Empty cell                                        ⚫ School

<u>Figure 1:</u> for each panel, the left side represents the residential map, and the right side the corresponding school assignment in the initial condition. Ties represent the connection between a household and the school to which it is originally assigned to.

Panel A displays an integrated map where parents are located at random. Panel B displays a "simple segregation" pattern where parents of each ethnicity no matter their tolerance level gather in very large ethnically homogenous clusters. Panel C displays a "complex segregation" pattern where intolerant parents of each ethnicity group self-select into large ethnically homogenous clusters, while tolerant parents of both ethnicities mix in areas in between these clusters.

Parents' decisions concern the choice of their child's school. Each household derives a utility value from sending their child to a particular school. The utility given by a school is derived from two components: the ethnic composition of the school, and the Euclidian distance between the household's residential location and the school's location.

Formally, the utility $U$ of school $j$, for agent $i$ of ethnic group $g$, is given by the following Cobb-Douglas utility function:

$$U_{i,g,j}(x_j, D_{i,j}) = S_g(x_j)^\alpha . D_{i,j}^{1-\alpha} \qquad (1)$$

where $\alpha$ is a parameter which determines the weight parents assign to the ethnic satisfaction component $S$ relative to the distance preference component $D$. The parameter $\alpha$ can vary between $O$ (only distance component matters) and 1 (only ethnic preference matters), and is the same for all members of the population, an assumption that we discuss further below.

The ethnic satisfaction $S$ depends on $x_j$, which represents the proportion of children of the same ethnic group $g$ as agent $i$, in school $j$. Following Zhang (2004), we adopt a functional form that allows to express a preference for more diverse schools (and neighborhoods) over more ethnically homogeneous ones. Technically $S$ is modelled as a single peaked step-wise linear function in $x_j$ (Zhang 2004). The ethnic satisfaction $S_g$ of school $j$ for an agent of ethnicity $g$ is given by equation (2):

$$S_g(x_j) = \begin{cases} \dfrac{x_j}{x_o}, & if \ x \le x_0 \\ M + \dfrac{(1-x_j)(1-M)}{(1-x_o)}, & if \ x > x_0 \end{cases} \qquad (2)$$

The symbol $x_j$ in equation (2) represents the proportion of agents of ethnicity $g$ in school $j$. When the proportion of in-group members $x$ is equal to its optimal value $x_o$, the function peaks, *i.e.* satisfaction reaches its maximum value 1. For $x \le x_o$, the satisfaction of the agent increases linearly with $x$. Above $x_o$, the satisfaction decreases linearly, until $x$ reaches 1 (100% of the school-population are in-group members) where the utility equals $M$. The parameters $x_o$ and $M$ characterize the ethnic preference structure of agents and are manipulated in our simulation experiments to model different levels of tolerance. The parameter $x_o$ expresses the proportion in-group members agents find optimal in a school, and the parameter $M$ captures how much satisfaction an agent derives from a school with 100% in-group members. When $x_o < 1$ and $M < 1$, agents prefer compositions with less than 100% in-group members above fully homogenous ones. The closer $x_o$ approaches $x_o = 0.5$ and $M$ approaches zero, the closer agents come to prefer perfectly mixed compositions above any other composition.

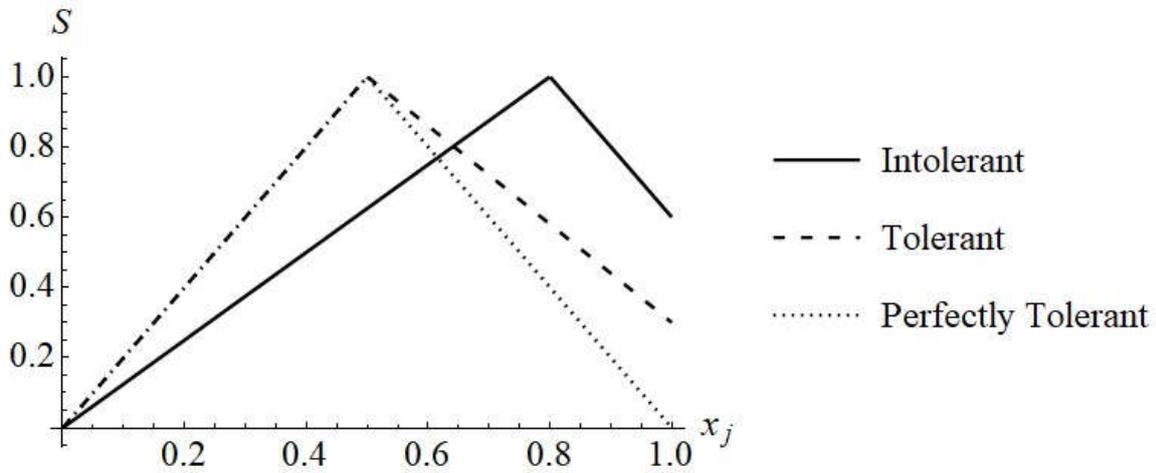

<u>Figure 2</u>: Single-peaked preference functions for intolerant ($x_o$ = 0.8, $M$= 0.6) vs. tolerant ($x_o$ = 0.5, $M$= 0.3) vs. perfectly tolerant ($x_o$ = 0.5, $M$= 0) parents, see equation (2).

Throughout our simulation experiments, we distinguish two types of tolerance. We model intolerant agents with $x_o = 0.8$ and $M = 0.6$, and tolerant agents with $x_o = 0.5$ and $M = 0.3$, with single-peaked preference functions as depicted in Figure 2. Reflecting empirical studies of residential ethnic preferences (e.g. Clark and Fosset 2008), this parametrization expresses the assumption that even intolerant agents do not favor perfect segregation but prefer presence of a small outgroup minority instead. Similarly, while tolerant agents reach their maximum utility when perfect integration is attained, they still have a relative preference for in-group members. For instance, they prefer 75% in-group members over 25%. Notice that both tolerant and intolerant parents still prefer schools with 100% co-ethnics to schools without any co-ethnics ($M > 0$). While this is empirically plausible, we also want to assess how this assumption affects model dynamics, using the limiting case of perfectly tolerant parent ($x_o = 0.5, M = 0$) as it also shown in Figure 2.

The other term of the utility function $U$, *i.e.* the benefit derived for agent $i$ from proximity of a school $j$, $D_{i,j}$, is given by equation (3), drawing on the formalization used by Stoica and Flache (2014).

$$D_{i,j} = \frac{dist_{max} - dist_{i,j}}{dist_{max}} \qquad (3)$$

In equation (3), $dist_{max}$ is used as a normalization factor and corresponds to the maximum distance between two cells on the torus map. $dist_{ij}$ is the distance between the agent's residential location and the school.

Dynamics of school choice are structured as follows. In every round of the simulation, 250 agents (approximately 4% of the population) are picked at random and decide in random sequence whether they will choose a new school, or remain in their current one. Each chosen agent computes her utility $U$ (equation (1)) for all $S = 30$ schools of the map including her current one, except in cases where some schools are full. We use probabilistic multinomial choice to model school choice (Bruch and Mare 2006; van de Rijt et al. 2009; Zhang 2004). Equation (4) formalizes the probability of a specific school $j$ to be chosen by agent $i$ within the set of $S$ available schools in the next time step $t$.

$$p_{ijt+1} = \frac{e^{\beta U_g(x_{jt}, D_{i,j})}}{\sum_{s=1}^{S} e^{\beta U_g(x_{st}, D_{i,s})}} \qquad (4)$$

Note that $\sum_{s=1}^{S} p_{ist+1} = 1$, in equation (4), where $p_{ijt+1}$ is the probability for agent $i$ of ethnic group $g$, to move to school $j$ at time $t + 1$ (the next round of the simulation); $\beta$ is a parameter that manipulates the weight of the utility $U$ relative to random chance (van de Rijt et al. 2009); $U_{i,j}(x_{jt}, D_{i,j})$ is given by equation (1). The multinomial choice model establishes an explicit connection between preferences of agents represented by the utilities $U$, and the decisions they make. The tightness of this link is modeled through $\beta$. As $\beta$ tends toward infinity, agents tend to behave as perfect utility maximizers. They rank the utilities associated with all available schools and (almost) always select the one ranked in the first place. Contrarily, as $\beta$ gets closer to zero, the probability of choosing a school depends less and less on its rank. In other words, $\beta$ allows to manipulate the strength of randomness in agents' choices. It can also be understood as a component that captures unspecified relevant characteristics in terms of attractiveness of schools, other than distance and ethnic composition. However, notice that $\beta$ is the same for all agents, thus there is no individual or group variability with regard to this random component of school choice. Throughout the simulation experiments in this paper, $\beta$ is fixed to a value of 12, as it is not a key parameter of interest. This value leaves room for a small degree of randomness in agents' choices which is enough for the model not to get stuck in instable equilibria. We checked whether our results critically depend on this specific value by replicating our experiments with $\beta = 100$ (see appendix D, figure D.1), and found no qualitative change.

Further, we let the school choice model run 140 rounds in each simulation run, giving on average every agent 6.08 opportunities to change school choice. As we will show below, this is in some conditions not long enough for the dynamics to approach a stable state. However, it seems highly implausible that pupil would change a school more than six times in their school career. We are therefore interested in the dynamics that unfold within this time frame but also explore dynamics for a much longer time frame in specific cases.

### 3.3. Map generator methods

To generate initial residential maps, we use a variant of a Schelling-Sakoda type algorithm very similar to our school-choice model. Details are given in appendix A. The residential segregation dynamic starts from a perfectly integrated map obtained by randomly allocating households to locations. We let the residential model run for different periods of time in order to create a continuum of maps. The longer the time period we simulate, the more segregated the map becomes. More precisely, we increment the number of rounds[1] in steps of 1, going from 0 to 70. The more chances agents get to relocate according to their ethnic preferences, the closer levels of residential segregation match the equilibrium segregation pattern corresponding to their preferences.

We distinguish two methods for creating residential maps, because the effect of tolerant parents on school segregation can be twofold: tolerant parents could reduce school segregation via more integrative school choices they make given their residential locations, but they could also reduce residential segregation via more integrative residential choices, which in turn will affect their residential locations and thus school segregation. In the following we describe how we separate the *direct* from the *indirect* effects of tolerant parents on school segregation by using two different types of residential maps.

---

[1] 1 round corresponds to 250 residential location decisions. A household can select either a new residential location from the set of available free locations to move into, or select the current location which means to stay.

### 3.3.1. Map generator method 1: "simple segregation" continuum

The first method we use assures that heterogeneity in ethnic preferences only affects school segregation through its *direct* effect on school choices. For this, varying degrees of residential segregation are obtained by artificially making tolerant parents behaving as if they were intolerant, but only when taking residential decisions. Put differently parents are all intolerant in the residential model but tolerant in the school model. With this method, we generate a continuum of maps that goes from non-segregated maps - that we call "integrated" maps (see panel A figure 1) to highly "segregated" maps, in which there are sharp boundaries between the two ethnic groups (see panel B figure 1). Although substantively less meaningful, this method ascertains that there are no systematic differences between tolerant and intolerant parents in terms of their residential locations. In other words, there is no segregation by tolerance even if there is ethnic segregation, this is why we label it "simple segregation".

### 3.3.2. Map generator method 2: "complex segregation" continuum

With the second map generating method, we let tolerant parents be tolerant also when taking residential decisions. This generates maps with both highly ethnically homogenous clusters *and* ethnically diverse areas. Additionally, it creates segregation by tolerance, that is, tolerant parents tend to live closer to tolerant parents from both ethnic groups, while intolerant parents self sort into highly homogenous ethnical clusters. Because these two forms of segregation co-exist in the same map, we label this phenomenon "complex segregation" (see panel c, figure 1)[2].

Notice that when gradually increasing segregation along a continuum of increasingly longer simulation runs, we start from "integrated" maps generated by assigning households uniformly randomly to locations on the map. However, the two methods generate qualitatively different residential patterns when segregation increases (see figure 1). As we let parents take more residential decisions, with the first method we get closer to the ideal-typical "simple segregation" pattern, while with the second method we approximate the ideal-typical "complex segregation" pattern, both depicted in figure 1.

---

[2] Because including tolerant parents' preferences in the residential decision making tends to reduce residential segregation levels, we slightly change the way we generate the continuum of maps for the purposes of method 2. In order to reach higher levels of residential segregation also with method 2, instead of incrementing number of rounds by one from 0 to 70 (as in method 1), residential segregation dynamics are run over a larger number of rounds up to a maximum of 90, sampling residential maps with varying levels of segregation for the first 60 rounds after every round and starting from round 63 up to round 90 after every third round.

## 3.4. Segregation indices

The key question we are interested in is how changing the proportion of tolerant parents in the population is related to the degree of school segregation, under different conditions of residential segregation and distance preferences. We thus need to measure both school segregation and residential segregation quantitatively. For this purpose, we adapt Massey and Denton's dissimilarity index (Massey and Denton 1988) for both. Equation (5) shows the general form of the index we use.

$$Dissimilarity = \frac{1}{2} \sum_{j=1}^{N} \left| \frac{g_{1j}}{G_{1T}} - \frac{g_{2j}}{G_{2T}} \right| \qquad (5)$$

where $g_{1j}$ and $g_{2j}$ are the number of agents of the first and second ethnicity respectively in local unit $j$. The unit $j$ can represents one of the $S$ schools (school dissimilarity index), or one of 256 tiles of 5x5 cells in our rectangular cellular grid (residential dissimilarity index), constructed such that each cell belongs to exactly one of the local units. $G_{1T}$ and $G_{2T}$ are the number of agents of ethnicity 1 and 2 respectively in the total population. Using the same index to measure both residential and school segregation allows to compare them.

Further, we are interested in two forms of segregation, segregation by ethnicity and segregation by tolerance. For both forms of segregation, we compute the same dissimilarity indices. Thus, for tolerance segregation $g_{1j}$ and $g_{2j}$ represent the number of tolerant and intolerant agents in local unit $j$, while $G_{1T}$ and $G_{2T}$ are the numbers of tolerant and intolerant agents in the total population. The higher the dissimilarity index, the more segregated a population is. Intuitively, the dissimilarity index measures the proportion of agents that would need to be relocated in a given distribution to obtain a perfectly integrated distribution in which the distribution in every local unit is a perfect representation of the distribution in the overall population.

## 4. Results

Our main interest is to understand how the proportion of tolerant agents affects school segregation in interaction with the variation in residential segregation and in strength of parents' preference for nearest schools. Our strategy is thus to fix all other parameters of the model, and vary only these three variables of interest and the two different types of residential maps shown in figure 1. Table 1 shows the baseline scenario defined by the fixed parameters in the model. These values are used in all simulation results displayed in the paper, except for some results where the changes in parameters are explicitly specified. The baseline scenario reflects a situation in which tolerant and intolerant parents clearly differ in the degree to which they prefer ethnically mixed schools or neighborhoods, but do not hold these preferences in an extreme way. Moreover, with $\beta = 12$, choices are largely determined by preferences but there is enough randomness to avoid fragile equilibria. All other parameters are chosen to represent a setting that can be deemed as prototypical for a small ethnically heterogeneous community with relatively small schools all of which are in reach for all households, but at varying travel distances. We consider this a setting suitable for our exploration of effects of ethnic tolerance and roughly resembling many empirical settings in which school choice occurs. The large ethnic heterogeneity and moderate size of schools and population in our baseline scenario create conditions in which we can expect parents' preferences to discernibly affect macro-level outcomes within a reasonable time frame. We modify some of these assumptions in robustness tests in this paper (see appendix D).

| Parameters common to both residential and school models: | |
|---|---|
| Number of parents-children (50% blue ethnicity, 50% yellow) | 5760 |
| $x_0$ intolerant parents | 0.8 |
| $M$ intolerant parents | 0.6 |
| $x_0$ tolerant parents | 0.5 |
| $M$ tolerant parents | 0.3 |
| $\beta$ | 12 |
| Percentage of empty cells | 10 |
| Number of parents taking a decision per round | 250 |

| Parameters specific to the school model: | |
|---|---|
| Number of schools | 30 |
| Number of rounds | 140 (*i.e.* 35 000 decisions in total; 6.08 decisions per parents on average). |
| School maximum capacity | 403 (*i.e.* 7% of the population) |
| $\alpha$ | [0:1] increment by 0.1 |
| **Parameters specific to the residential model:** | |
| Number cells considered including current | 9 |
| Radius of the Neighbourhood | 6 |
| Number of rounds method 1 | [0:70] increment by 1 |
| Number of rounds method 2 | [0:60] increment by 1; [63:90] increment by 3 |
| **<u>Table 1</u>: Values and range of variation of parameters.** | |

In what follows, we first present simulation experiments that use the first method for generating the residential map. In these experiments, diversity in tolerance affects school segregation only through parents' school choices. In a second set of experiments we relax this assumption and use method 2 to generate maps that display "complex segregation". We investigate how diversity in tolerance affects school segregation both through its effects on the residential locations parents choose – *indirect effect* - and through their school choices given the "complex segregation" pattern – *direct effect*. In all experiments, our main focus is on how school segregation is affected by the complex interplay of parents' ethnic tolerance, the residential pattern and parents' distance preferences.

## 4.1. Experiments with method 1 residential segregation

### 4.1.1. Baseline experiment: all parents are equally intolerant

We start with a baseline experiment in which all agents are intolerant. Figure 3 shows how for this condition residential segregation is associated with school segregation for varying strength of parents' preference for nearby schools (1-$\alpha$). Figure 3 is drawn based on 2343 data points: 11 values of $\alpha$, 71 values of rounds for the residential model corresponding to 71 different levels of segregation, and 3 independent realization for each combination of parameters. We compute the value of the residential dissimilarity index RDI (x-axis) and the school dissimilarity index (z-axis) at the end of each simulation run. We split RDI into intervals of width 0.05 and average all values of SDI that correspond to each combination of $\alpha$ and RDI.

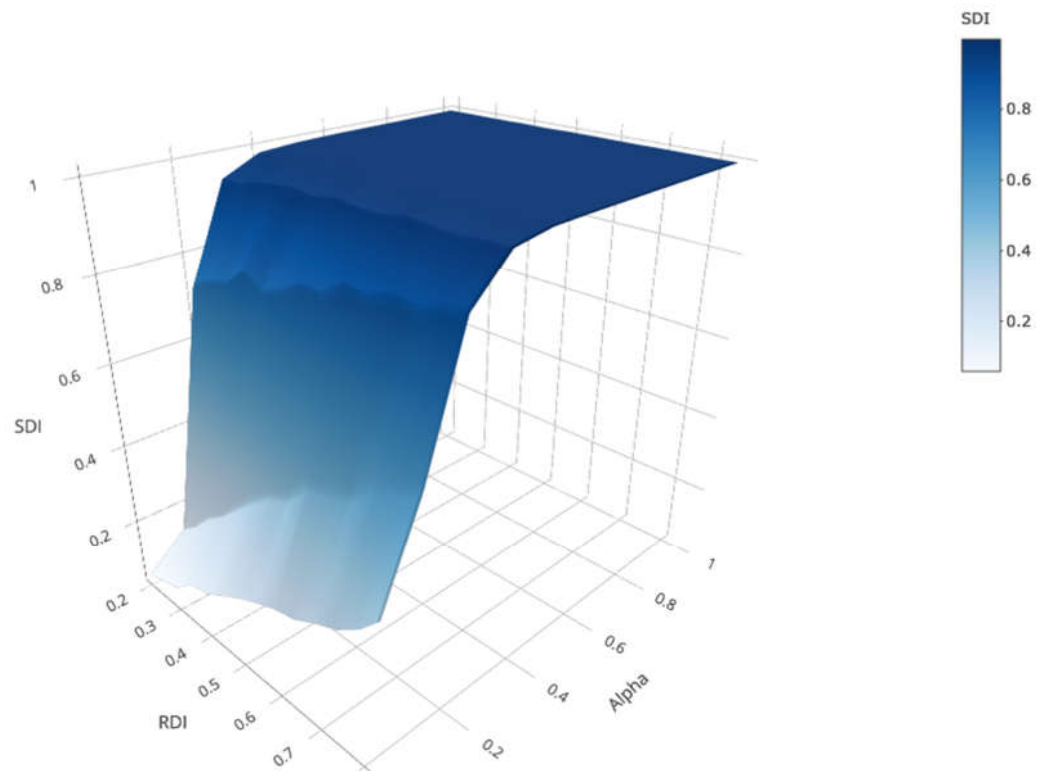

<u>Figure 3</u>: Relations between residential dissimilarity index RDI (x-axis), weight of ethnic preference $\alpha$ (y-axis) and school dissimilarity index SDI (z-axis), for 0% tolerant agents, for the "segregated" maps continuum generated with map generator method 1. The surface is drawn based on 2343 observations each corresponding to one complete simulation run outcome. Since RDI is continuous we split it into intervals of width 0.05. We average all three values of SDI that correspond to the same combination of $\alpha$ and RDI.
https://plot.ly/~lsage/33

As figure 3 shows, in the scenario where agents evaluate schools only in terms of distance $(\alpha = 0)$[3] school segregation increases linearly with residential segregation. The reason is that schools' ethnic compositions mirror their catchment areas' compositions. Agents rarely change schools because the least distant school they initially attend maximizes their utility. Moves are only due to randomness.

---

[3] It is interesting to note that even for very high levels of residential segregation (RDI=0,85), when $\alpha = 0$ school segregation remains moderate (SDI=0,42). Two phenomena explain this outcome. Firstly, schools are positioned at random on the map, which leads some of them to be located close to frontiers between residentially segregated areas. Their catchment areas contain children from both sides of the residential ethnic border, therefore their ethnic composition is relatively integrated, especially for the bigger ones. This brings down the overall initial level of school segregation. Secondly, detailed analyses of simulation runs revealed that for highly residentially segregated maps, the level of school segregation further decreases over time in a simulation run. Recall that under $\alpha = 0$ tolerant and intolerant parents are equivalent because they only minimize distance

This close link between residential segregation and school segregation starts to change as soon as agents consider the ethnic mix in schools ($\alpha > 0$). From $\alpha = 0.1$, the level of school segregation (SDI) increases for all levels of residential segregation as parents put increasing weight on schools' ethnic composition. When $\alpha$ reaches $0.2$, there is a take-off where the model starts generating noticeably higher levels of school segregation, so that even on maps with low levels of residential segregation, schools end up being highly segregated. The surface gets almost perfectly flat for $\alpha = 0.3$ and above, showing that the moderating effect of low levels of residential segregation on school segregation almost completely disappears when distance constraints become weaker. For these values of $\alpha$, school segregation exhibits the self-reinforcing dynamic well known from the Schelling-Sakoda model. High levels of school segregation emerge although parents' preferences could be satisfied with more diverse schools.

To sum up, when all agents are intolerant, the stronger the distance constraint, the more school and residential segregations align and self-reinforcing school segregation is curbed by distance preferences. While this reflects insights from earlier research (Stoica and Flache), our results also show how distributions with integrated schools become less stable when it is considered that school choice has a random component. Using an otherwise comparable model with deterministic preference functions, Stoica and Flache (2014) found that initially integrated compositions remained stable at considerably higher levels of $\alpha$. The reason is that our multinomial choice model generates more school changes also in otherwise "frozen states" so that school choices more easily trigger the preference dynamics leading to self-reinforcing segregation.

### 4.1.2. Introducing heterogeneity in ethnic preferences: 50% of tolerant agents

---

from home to school. However, this minimization is not perfect as $\beta = 12$ allows for some randomness in decisions, weakening the association between residential map characteristics and school segregation. If agents' choices would perfectly reflect their preferences – in this case staying in the closest school from their home – the level of school segregation would be higher. For example, for $\beta=100$ and RDI=0,85, we obtain SCI=0,55 (instead of 0,42 for $\beta=12$), all other things being equal.

In this section, we assume that in both ethnic groups 50% of the population is tolerant. Results will be compared to what we obtained with 100% intolerant parents. Our main outcome of interest is the difference that the presence of tolerant agents makes for school segregation. Figure 4 displays results in the same way as figure 3, but now based on simulations with 50% tolerant households. Figure 5 highlights how adding 50% tolerant parents changes school segregation outcomes under different conditions, charting the difference in levels of SDI generated by the model with 50% and with 0% tolerant agents across all point of the parameter space (5a: 3-D plot, 5b: contour lines plot).

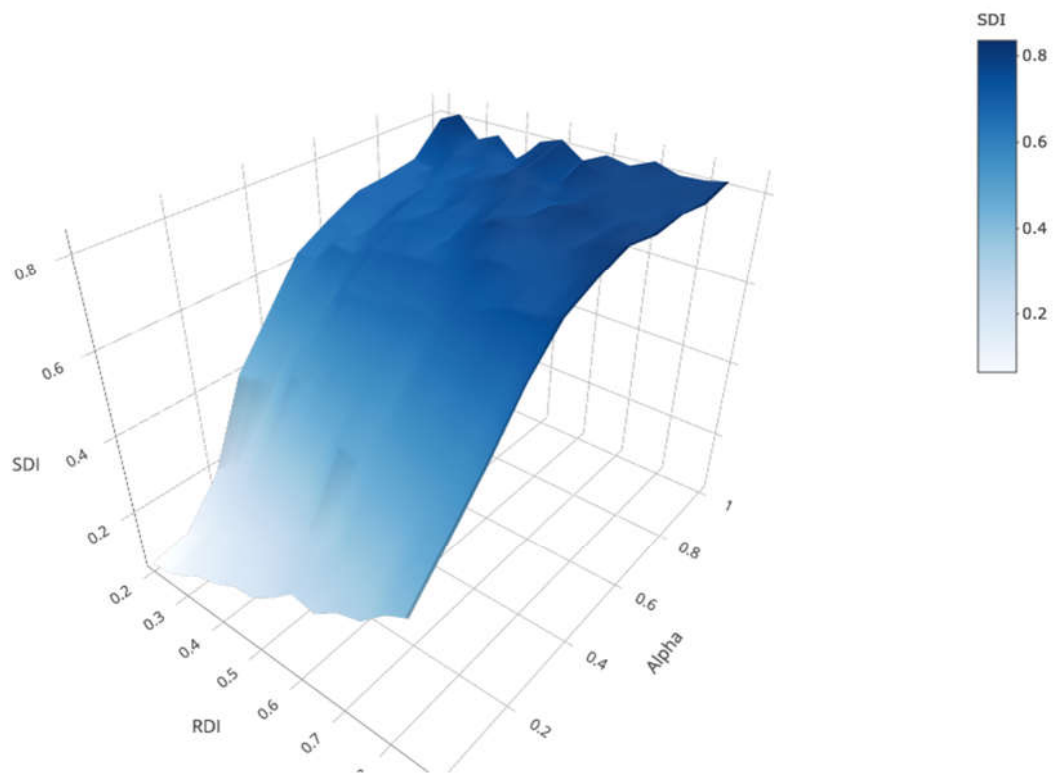

Figure 4: Relations between residential dissimilarity index RDI (x-axis), weight of ethnic preference $\alpha$ (y-axis) and school dissimilarity index SDI (z-axis), for 50% tolerant agents, for the "segregated" maps continuum generated with map generator method 1. The surface is drawn based on 2343 observations each corresponding to one complete simulation run outcome. Since RDI is continuous we split it into intervals of width 0.05. We average all three values of SDI that correspond to the same combination of $\alpha$ and RDI.
https://plot.ly/~lsage/31/#/

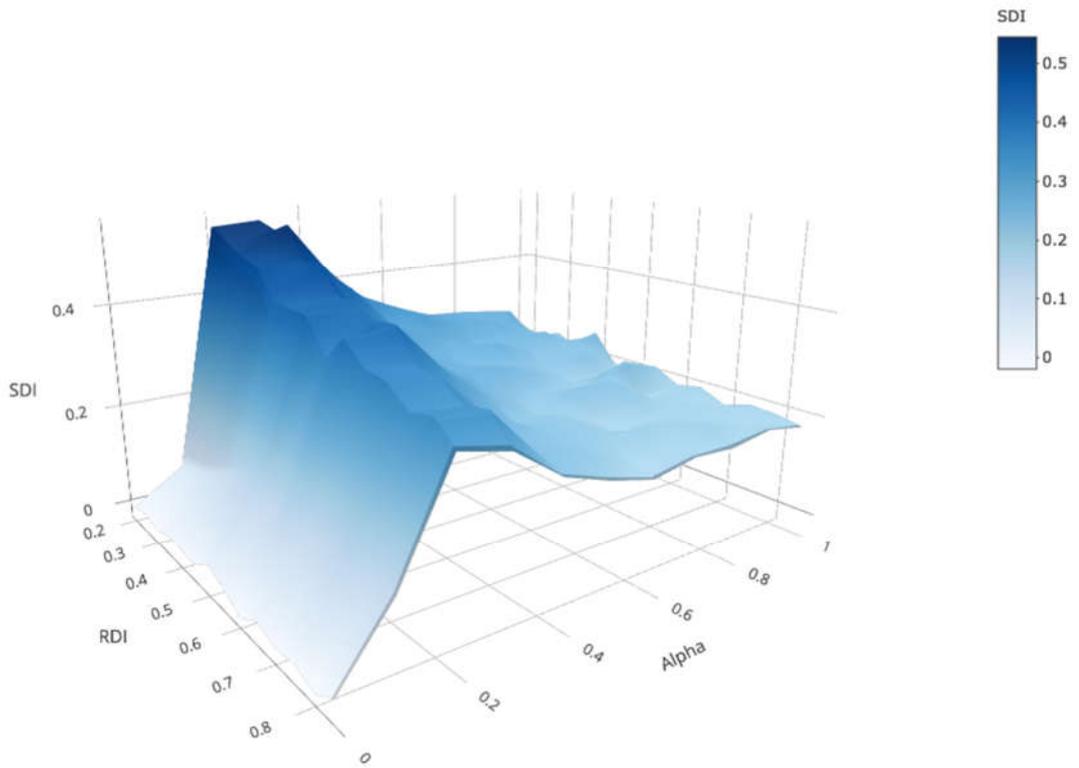

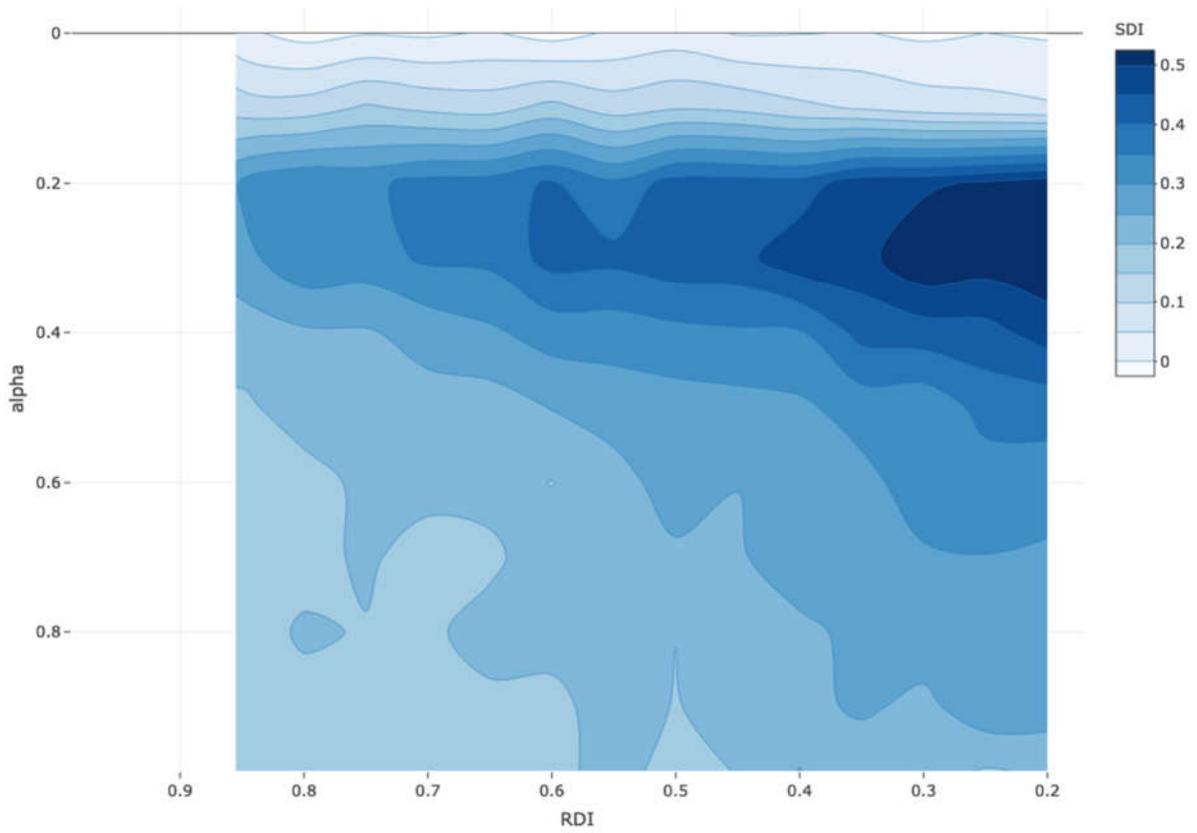

Comparison with the baseline condition without tolerant agents shows two major results. First, also with 50% tolerant agents we observe the desegregating effect of distance preference on school segregation: the stronger the preference for nearby schools (lower $\alpha$), the lower the school segregation for all levels of residential segregation, ceteris paribus. Second, for all values of $\alpha > 0$, the introduction of tolerant agents decreases school segregation as compared to the baseline experiment.

However, there are clear differences in the extent to which the presence of tolerant agents reduces school segregation (figure 5). Two main patterns can be discerned. First, we find that the presence of tolerant agents reduces school segregation noticeably only when distance constraints are neither too strong nor too weak. With strong distance preferences, below $\alpha = 0.2$ parents strictly prefer nearby schools to more distant schools, regardless of their ethnic tolerance. At the other extreme, with weak distance preferences ($\alpha > 0.6$), the presence of tolerant agents only moderately dampens the self-reinforcing dynamics of school segregation that we also found when all parents were intolerant. As figure 5 shows for $\alpha = 1$, school segregation with 50% tolerant parents falls only moderately below the maximal degree of segregation generated with only intolerant parents, across all residential maps (around 0.19 difference in SDI). The second pattern exhibited by figure 5 is that the degree to which tolerant agents reduce school segregation also depends on the level of residential segregation. Across all conditions, but most visibly for $0.2 \leq \alpha \leq 0.6$, we found that the more integrated the residential map, the more the presence of tolerant agents reduces school segregation. All in all, according to our results the "sweet spot" in the parameter space where the presence of tolerant parents has the biggest impact on reducing school segregation is characterized by moderate preferences for nearby schools and high residential integration.

Two assumptions of our model turn out to be of particular importance for understanding these results, both representing aspects of school choice dynamics in the real world. The first assumption is that even tolerant parents are only 'moderately tolerant'. The second important assumption is that schools have a maximum capacity to receive new pupils.

As figure 2 shows, the preference curves of tolerant and intolerant agents are not perfectly symmetrical. Tolerant parents favor perfectly integrated schools in our model, but when faced with the choice between being in a clear minority position or a clear majority position, they prefer the latter[4]. To see how asymmetry of the ethnic preferences of tolerant parents contributes to generate the inverted U-shaped effect that tolerant parents have on school segregation (see figure 5), consider first an illustrative scenario in which the residential map is strongly integrated and parents are indifferent regarding home-school distance at all ($RDI < 0.25$ and $\alpha = 1$). At the outset, all schools are highly integrated in this situation, with approximately a 50/50 mix of ethnicities. In the presence of 50% tolerant agents only the members of the intolerant half of the population are clearly dissatisfied with this situation. Although these parents are not constrained by a distance preference, they find initially no alternative schools which are clearly more attractive than their current school. Yet, due to small random variation in the initial composition of schools, intolerant agents can slightly improve their utility by migrating to schools with a somewhat higher proportion of in-group members. This increases the chances that further intolerant parents of the same ethnic group follow suit, gradually 'tipping' their target schools towards hosting a majority of the initially overrepresented group. As this process happens simultaneously in all schools, the intolerant half of the population begins to segregate into ethnically increasingly homogenous schools due to self-reinforcing preference dynamics like those known from the Schelling-Sakoda model. At the same time, in this phase of the dynamics, tolerant parents stay behind in the mixed schools they were satisfied with from the beginning, or they move out of emergent ethnically tipping schools into free locations in integrated schools abandoned by intolerant parents. As a result, we observe increasing segregation of the parent population by tolerance as well as by ethnicity. Figure 6 plots the evolution of ethnic and tolerance segregation in schools over an illustrative run in this condition. As the upper part of figure 6 shows, in the first 50 rounds both indices quickly increase, which indicates that tolerant parents cluster into integrated schools, while intolerant parents tend to separate both from the ethnic out-group and from tolerant parents in their ethnic in-group. The bottom part of the figure demonstrates how in a ceteris-paribus replication without tolerant parents, ethnic segregation rapidly and steadily reaches perfect segregation in this scenario. This illustrates how ethnic segregation is curbed by the presence of tolerant parents.

---

[4] Technically, the slope of satisfaction with the ethnic composition is steeper in deviations from the 50% optimal mix on the left part of the optimal point ($x \leq x_0$), than on the right side ($x > x_0$).

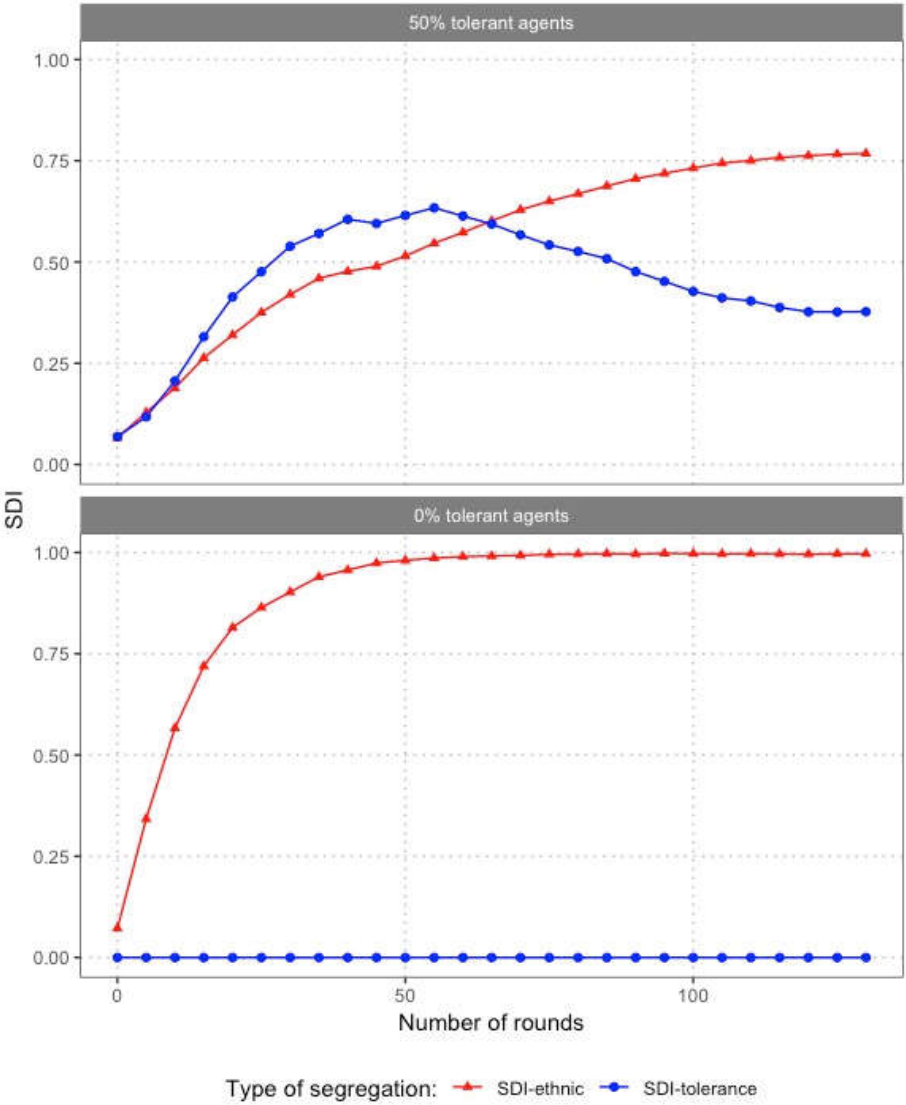

: Change of ethnic school segregation (SDI-ethic) and tolerance school segregation (SDI-tolerance) within the first 140 rounds of illustrative runs. Upper panel: 50% tolerant parents (all intolerant). Lower panel: 0% tolerant parents. $RDI < 0.25, \alpha = 1$.

Yet, as figure 6 (upper panel) further demonstrates, the mixed schools populated by tolerant parents are not stable. After about 60 rounds, tolerance segregation starts to plummet while ethnic segregation continues to rise. The key reason for this unravelling of integrated schools is the asymmetric ethnic preference of tolerant parents. While tolerant parents favor schools with a 50/50 mix, random movements as well as the uncoordinated outflow of intolerant parents entail random variation of the ethnic school composition around this optimal point. Consequently, tolerant parents of the ethnic group that happens to become a small minority in their school at some point, face the choice between their current school and alternative schools in which their group is overrepresented. Beyond some critical difference, they prefer the alternative schools with an overrepresentation of their ethnic in-group.

We suspected this progressive tipping of integrated schools had no reason to stop at 140 rounds, since it mainly depends on randomness, that is, on a few parents of one group who consecutively move out of an integrated school, gradually shifting the ratio away from the 50-50 mix, which locally triggers a segregating cascade. To further test this possibility, we ran additional 50 replications of 1400 rounds each. Confirming our intuition, simulations revealed that a second phase of a self-reinforcing preference dynamic was triggered sooner or later, this time among tolerant parents. In this phase, ethnic segregation of schools increases to a level that eventually approximates that of a world without tolerant agents, while segregation by tolerance erratically declines each time a mixed school tips. On average SDI reached 0.86 which indicates that without any distance constraint, in the longer run, the desegregating effect of tolerant parents is considerably lower than suggested by figure 5. Interestingly, this reflects results of recent empirical research showing the existence of tipping behavior of schools (Caetano and Maheshri 2017; Spaiser et al. 2018). Appendix B displays the results of one illustrative run over 1400 rounds, demonstrating the progressive unravelling of initially integrated schools.

The asymmetric ethnic preferences of tolerant parents made integrated schools unstable in our illustrative scenario without a preference for nearby schools ($RDI < 0.25$, $\alpha = 1$). With stronger preference for nearby schools, dynamics change fundamentally. Consider the average ethnic school segregation figure 7 shows for $\alpha = 0.3$ in residentially integrated worlds (low RDI) for comparison. Even after 1400 rounds ethnic school segregation was stable and on average over 50 runs we found $SDI = 0.58$ ($sd = 0.04$), well below the level of almost $SDI = 0.98$ ($sd = 0.004$) without tolerant parents. Figure 7 demonstrates this with a comparison of two illustrative runs with and without tolerant parents for this condition.

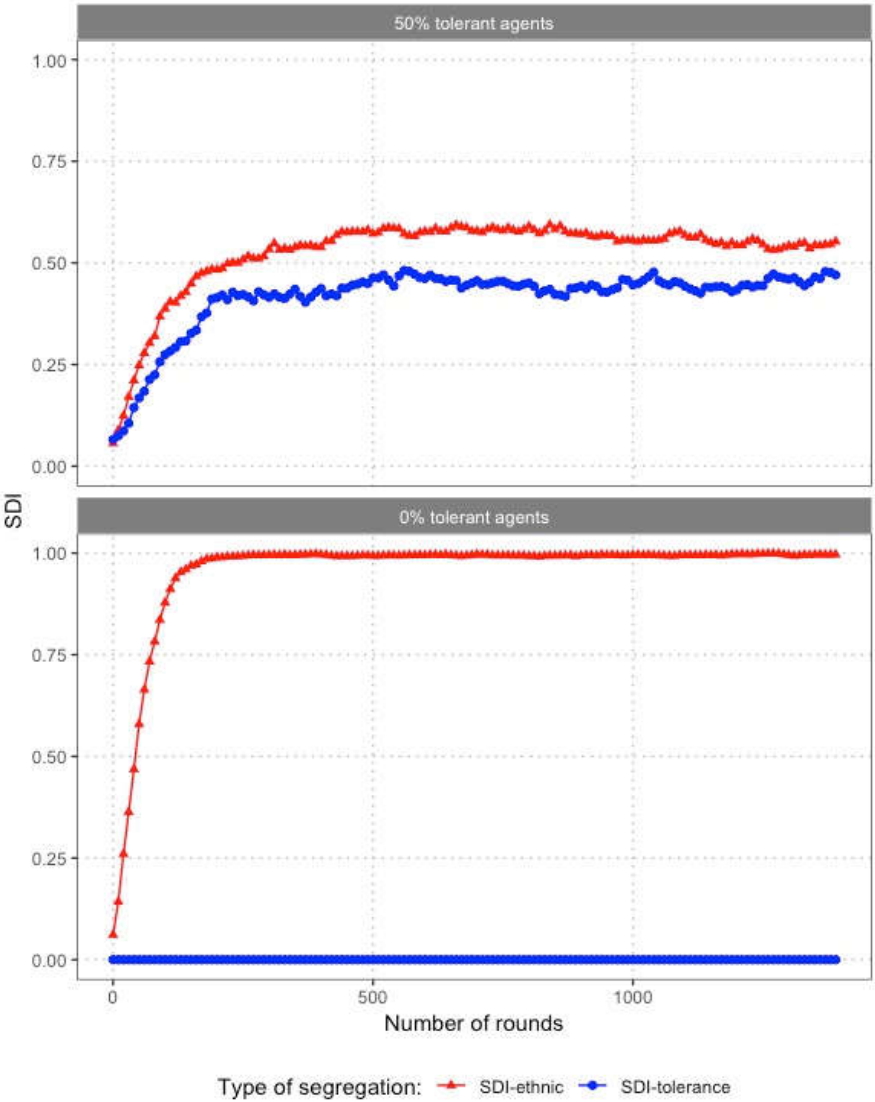

Figure 7: Change of ethnic school segregation (SDI-ethic) and tolerance school segregation (SDI-tolerance) within the first 1400 rounds of illustrative runs. $RDI < 0.25$, $\alpha = 0.3$. Upper panel: 50% tolerant parents. Lower panel: 0% tolerant parents (all intolerant).

Based on figures 5 and 7 we can tentatively conclude that in a residentially integrated setting, the combination of moderate distance preferences and the presence of tolerant parents can reduce school segregation. Relocating from an initial school is more costly if there is a distance constraint, thus parents are only likely to do so if they find an alternative school that is considerably more attractive to them in terms of ethnic composition. In integrated maps, such schools are rare in the initial situation. And even when some mixed schools tip, it requires much stronger random fluctuations for the remaining mixed schools to also tip. The reason is that the tolerant parents who had previously placed their children in the schools that eventually tipped, increase the demand for the remaining ones, such that it becomes less and less likely that any slight deviation from the 50-50 mix will trigger a cascade that destabilizes the integrated school. As our results show, this effect of the costs of choosing distant schools is just enough to dampen ethnic preference dynamics in this region of the parameter space. Yet, some movement still occurred primarily by dissatisfied intolerant parents, so that ethnic and tolerance segregations increased above their initial level but then stabilized at an average of about 0.55 for SDI-ethnic and 0.45 for SDI-tolerance. Thus, ethnic segregation in schools is comparable to what we found to be only a temporarily stable level without distance constraints, while tolerance segregation is somewhat lower.

The combined effect of moderately high distance constraints and the presence of 'moderately tolerant' parents that we observed for residentially integrated maps occurs qualitatively at all levels of residential segregation inspected in this simulation experiment (see figure 5). At the same time, figure 5 reveals a consistent pattern of smaller differences between worlds with and without tolerant agents when the initial residential map is segregated. The reason is that in highly segregated residential maps, the potential dynamic of segregation by tolerance is limited by the asymmetric ethnic preferences of tolerant parents, in combination with two further factors: first, the number of schools that are candidate to being integrated and second, their limited capacities to host pupils. In a residentially segregated map, only few schools are initially ethnically integrated. These are schools located close to boundaries between ethnic clusters in the residential map. While intolerant parents of both ethnicities will tend to get away from these few ethnically diverse schools, tolerant parents on both sides of the borders will instead be attracted by them. Yet, once these schools are filled, tolerant parents who cannot find a place in such schools must choose between less satisfactory options. Due to their asymmetric ethnic preferences, they prefer to stay in ethnically homogenous schools of their own group over schools where they are clearly in the minority. When distance preferences comes into play, this process is further amplified by the fact that many tolerant parents reside inside of ethnic clusters. For them, mixed schools on the boundaries between clusters are prohibitively distant. Additionally, it also is almost impossible that any school that is initially segregated can desegregate. The reason is that an initially segregated school remains highly attractive for intolerant parents, and non-attractive for tolerant parents of the other group. It would therefore require a long chain of unlikely decisions to desegregate a school.

To sum up, in a residentially segregated map tolerant parents are more limited than in an integrated map to choose integrated schools. The interplay of these factors explains why in figure 5 for all values of $\alpha$ the effect of tolerant parents is stronger in residentially integrated settings. Additionally, we found that it is the combination of asymmetric preferences and limited school capacity that moderates the effects of tolerant parents on school segregation particularly in residentially segregated settings. As a further test, we simulated scenarios in which both conditions were changed. We assumed an unlimited number of students per school, and set $M = 0$ for tolerant parents, making their preference function symmetric (see figure 1 and appendix D, figure D.3). Figure 8 shows a comparison of illustrative runs with and without these modifications for the setting of a highly segregated residential map and a moderate distance constraint of $\alpha = 0.4$, with a 50% share of tolerant agents.

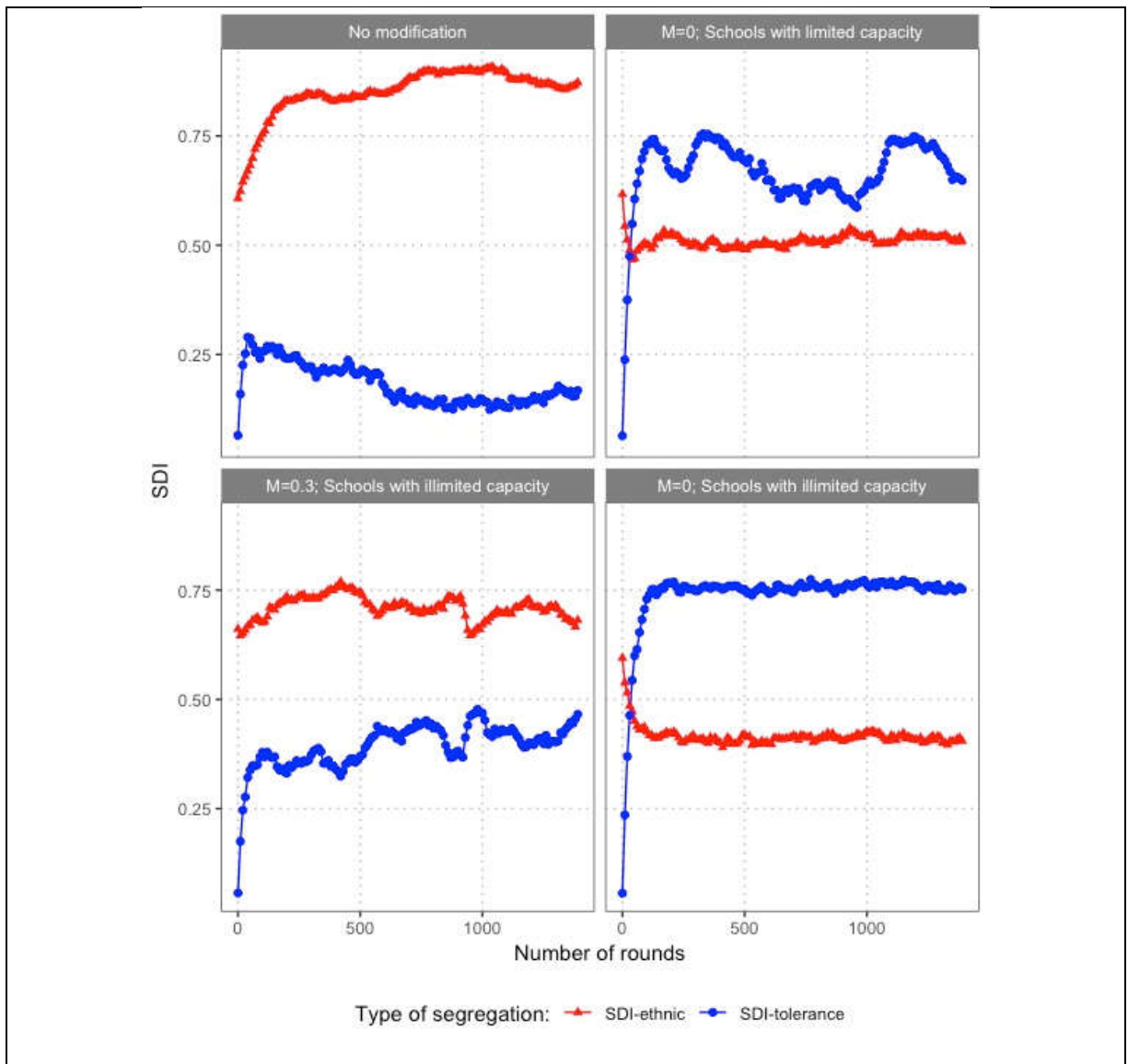

<u>Figure 8</u>: Evolution of school segregation indices over one simulation run of 1400 rounds starting from highly segregated maps ($RDI > 0.8$, "simple segregation" map generator method 1); $\alpha = 0.4$; $x_0 = 0.5$ (the peak of the ethnic satisfaction function) for the 50% tolerant parents (unchanged). The upper left panel corresponds to the standard parametrization of the model with schools' maximum capacity equal to 403 pupils, and $M = 0.3$ for tolerant parents; the upper right panel corresponds to the standard parametrization except that M=0 for tolerant parents; the lower left panel corresponds to standard parametrization except that schools have no limits to their welcoming capacity; the lower right panel correspond to the standard parametrization except M=0 for tolerant parents and schools have no limits to their welcoming capacity.

Figure 8 further confirms our explanation. We see significantly higher levels of segregation by tolerance in schools for all modified parametrizations. The modifications reduce at the same time ethnic school segregation because segregation by tolerance prevents mixed schools from unravelling due to preference cascades among tolerant agents.

Figure 9 charts the combined effects of distance constraint and residential segregation on school segregation, broken down by tolerance-conditions. Simply put, a weaker preference for nearby schools simultaneously increases segregation by ethnicity, and reduces segregation by tolerance in schools.

The second result is considerable variance between runs with similar parametrization as parents' preference for nearby schools ($\alpha$) becomes weaker. With a weak distance constraint, the movements of intolerant parents lead initially to strong segregation of schools by tolerance, but beyond some tipping point several of the remaining mixed schools also tend to unravel and become ethnically more homogenous[5]. When exactly the tipping point occurs depends on when each mixed school will tip, which in turn depends on a number of random factors such as the sequence in which agents are activated, or the distance between schools. This explains the variation in tolerance segregation depicted in the lower panel of figure 9. Tolerance segregation is high in realizations that have not yet reached the point where many schools have tipped when the simulation ends. Instead, it is considerably lower when dynamics have passedthis point. In some cases, integrated schools even unravel entirely into segregated schools. Overall, the weaker the distance constraint the higher the standard deviation of tolerance segregation between replications, reflecting more instability of the tolerance segregation that develops early on in the process. How stable is this result in the longer run? In long-term simulations (rounds=1400) we find that for $\alpha = 1$, SDI-tolerance diminishes to 0.27 for RDI=0.25 and the standard deviation equals 0.15. For RDI=0.67 we find that SDI-tolerance=0.24 and std=0.12 (see table 1) after 1400 rounds.

---

[5] We use the word tipping both at the global level, *i.e.* the simulation run tips, and at the local level, *i.e.* a specific school tips. However, the former is just the aggregated consequence of an accumulation of the latter.

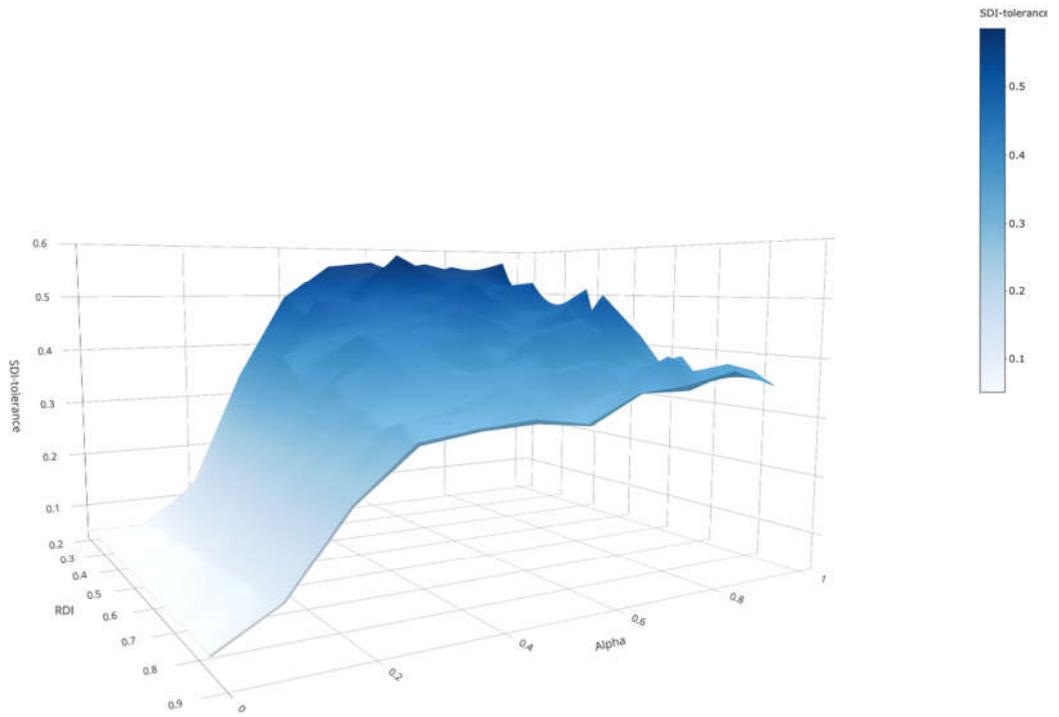

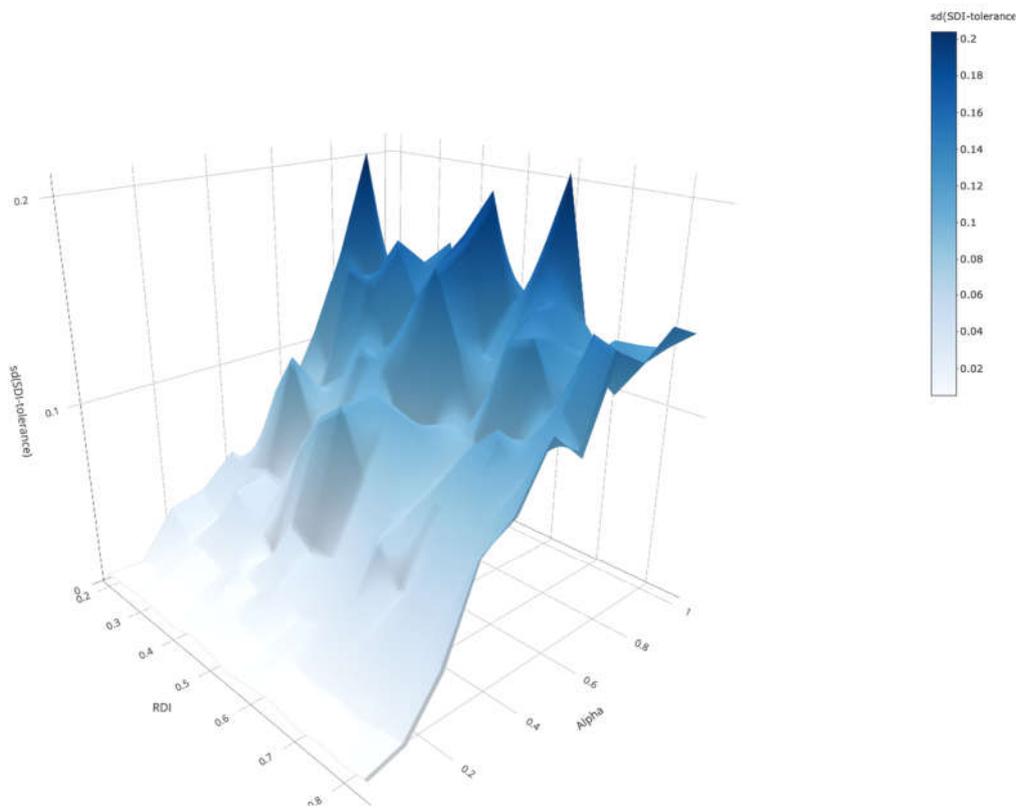

Figure 9: Relations between residential dissimilarity index RDI (x-axis), weight of ethnic preference $\alpha$ (y-axis) and school dissimilarity index by tolerance SDI-tolerance (z-axis), for 50% tolerant agents, for the "simple segregation" continuum generated with map generator method 1. The surface is drawn based on 2343 observations each corresponding to one complete simulation run outcome. Since RDI is continuous we split it into intervals of width 0.05. Upper panel: we average all three values of SDI-tolerance that correspond to the same combination of $\alpha$ and RDI. Lower panel: standard deviation of the corresponding mean of SDI-tolerance.

https://plot.ly/~lsage/39
https://plot.ly/~lsage/41

## 4.2.    From "simple" to "complex" segregation

In this section, we examine how tolerant parents affect school segregation when parents' residential location is not independent from the preferences that underlie their school choice behavior. We now use map generator method 2, creating a continuum of increasingly segregated maps going from "integrated" to "complex" segregation (see figure 1). In "complex" segregation, tolerant parents tend to reside in ethnically integrated residential areas, while less tolerant parents choose homogeneous residential neighborhoods, reflecting their ethnic preferences.

Figure 10 plots the results of a ceteris paribus replication of experiment 1, showing the magnitude of change in the school segregation index when outcomes with 50% tolerant parents are compared to outcomes with 0% tolerant parents (as in figure 5), while residential maps were generated by method 2 instead of method 1.

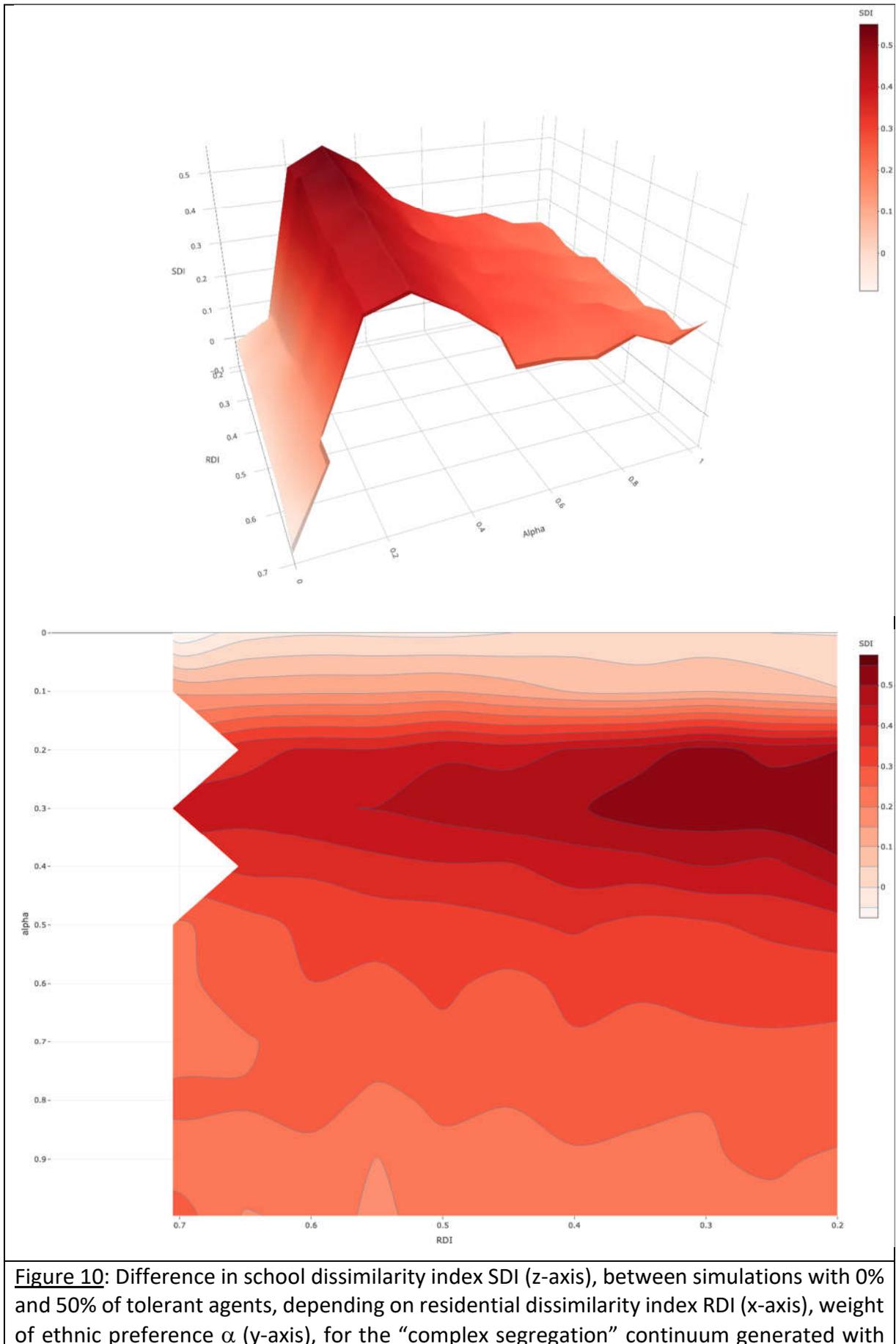

Figure 10: Difference in school dissimilarity index SDI (z-axis), between simulations with 0% and 50% of tolerant agents, depending on residential dissimilarity index RDI (x-axis), weight of ethnic preference $\alpha$ (y-axis), for the "complex segregation" continuum generated with map generator method 2. The surface is drawn based on 4686 observations each corresponding to one complete simulation run outcome. Since RDI is continuous we split it into intervals of width 0.05. We average all values of SDI that correspond to the same combination of $\alpha$ and RDI separately for 0% and 50% of tolerant agents. We then compute the difference between the values obtained in each case. Example: for $RDI = 0.2$ and $\alpha = 0.2$, the 50% tolerant agents reduce average SDI by 0.53.
https://plot.ly/~lsage/43
https://plot.ly/~lsage/45

Figure 10 shows that the qualitative effects of tolerant parents on school segregation observed with map generator method 1 are largely replicated with method 2. The inverted u-shaped effect of the distance preference alpha on how tolerant parents affect school segregation is also obtained with map generator method 2. In the region $\alpha \geq 0.3, RDI > 0.4$, we observe even lower ethnic school segregation in "complex" maps than in "simple" maps. This suggests that if anything, the inverted u-shaped effect is more pronounced with complex maps.

However, there are also some noticeable differences between the results we obtain for the two map generator methods. The second qualitative effect observed in experiment 1 was that tolerant parents have a lower desegregating effect when residential segregation is high. This result is much weaker with the "complex" maps continuum. For illustration, under method 2 moving from the lowest to the highest level of residential segregation reduced the effect on average school segregation in terms of $SDI$ by about 25% at $\alpha = 0.3$, where tolerant parents had the biggest impact. With method 1 the corresponding difference reduction at $\alpha = 0.3$ was much more substantial, almost 50%.

Notice that for high residential segregation the two methods cannot straightforwardly be compared because the map generator method 2 hardly generates $RDI > 0.7$ so that the difference between the lowest and the highest level of residential segregation is higher for method 1. The reason is the sorting of tolerant and intolerant in different residential areas. Similarly to the previous analyses for school segregation dynamic, this time, the residential model generates separate clusters of intolerant agents of the same ethnicity and ethnically diverse clusters of tolerant agents that reduce the level of residential segregation, which is in line with (Hatna and Benenson 2015). This is important to consider when we compare the effects of tolerant parents across the two kind of maps. Since the differences between figure 5 and 9 (appendix C) are very small meaning that the *direct effect* of tolerant parent is almost the same in simple and complex segregation, yet a bit more pronounced for higher levels of residential segregation. However, heterogeneity in ethnic preferences does not only reduce school segregation for similar levels of residential segregation, but it also significantly reduces residential segregation in the first place. Thus, we find support for an *indirect effect* of heterogeneity in tolerance on school segregation.

Finally, we wanted to know whether complex segregation patterns make the effect of tolerant parents more stable in the long run. In the discussion of experiment 1 we found that in some conditions, mixed schools which initially formed due to school choices made by tolerant parents turned out to not be stable in the long run. Intuitively, we might expect that in "complex" maps this is different. Schools located in ethnically diverse areas populated by tolerant parents should be more stable, precisely because the tolerant parents are located close to them. Therefore, distance preference should even reinforce the attractiveness of mixed schools in the eyes of tolerant parents. To test this intuition, we simulated 100 runs of 1400 rounds with each of the two map generator methods for a scenario where a clear difference between segregation dynamics in the two types of maps can be expected to be arise.

We chose moderate strength of preferences for nearby schools, $\alpha = 0.3$, because this is the point where the desegregating effect of tolerant parents is strongest. Further we selected a relatively high degree of ethnic residential segregation, $RDI = 0.67$, because here differences between maps with tolerant parents selected into mixed neighborhoods and maps with ethnic clustering unrelated to tolerance preferences can be expected to show up clearly. In fact, $RDI = 0.67$, is almost the maximum residential segregation we can obtain for "complex" maps with 50% tolerant agents. Figure 11 plots the results of this experiment. In line with our intuition, "complex" segregation slightly reduces ethnic segregation in schools (after 1400 rounds: $\Delta SDI = -0.056$, i.e. 8.2% decrease). We also observe a substantial rise in segregation by tolerance in schools which also fits with our interpretation. Yet, one can see that the dispersion between runs remains important, even in the long run. This is likely due to different factors, in particular we believe two aspects play a role here: the relative position of the schools, and the order in which agents are activated during the run in particular at the beginning. To find out how exactly each of these factors relates to the variation around the mean between different runs under the same conditions would require further investigations which lead beyond the current paper.

Finally, the average dynamics appear to be different across the two types of maps. While on the simple maps ethnic segregation keeps increasing gradually throughout the 1400 rounds of our simulation, it seems to have stabilized after about 500 rounds on the complex maps. Conversely, tolerance segregation is still increasing slightly on the complex maps after 1400 rounds but seems to have stablized on the simple maps. This is further in line with the intuition that the self-sorting of tolerant agents into mixed residential areas sustains mixed schools in those areas in the long-run, while mixed schools tend to slowly unravel on simple maps.

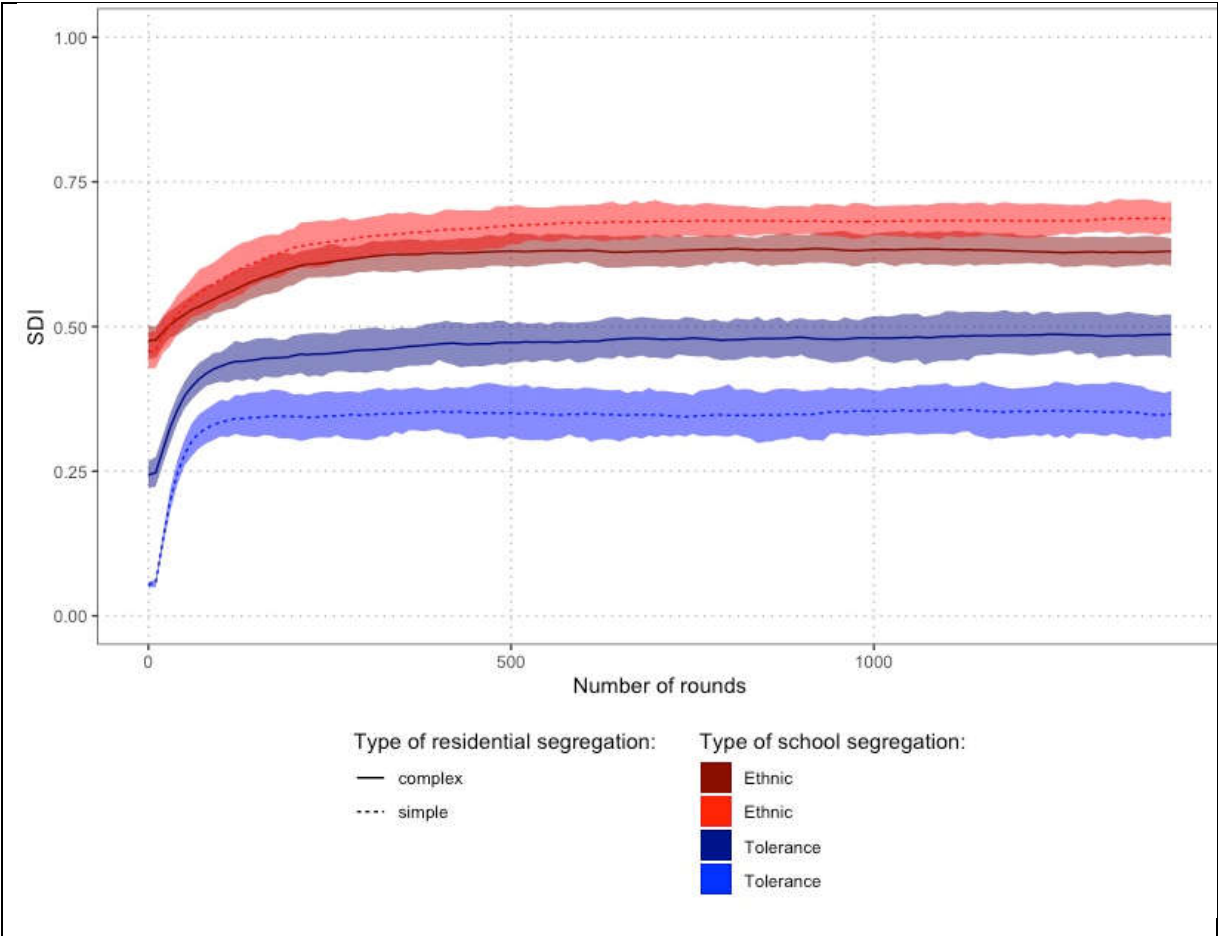

Figure 11: Evolution of school segregation indices over 200 simulation runs of 1400 rounds each. Initial conditions: $RDI = 0.67$, $\alpha = 0.3$. 100 runs started on "simple" residential maps generated with method 1 and 100 runs started on "complex" residential maps generated with method 2. Lines correpond to the average of the SDI-ethnic and of the SDI-tolerance over the 100 runs. The interval around the average encompasses 75% of the observations (quartile 1 and 3).

**Discussion**

In this paper, we addressed the question whether and how a large fraction of tolerant parents could reduce school segregation. We developed a new model of school choice that allows to assess the potential for self-reinforcing preference dynamics in school segregation similar to those identified by the Schelling-Sakoda model of residential segregation. For this, we modelled school choice based on a preference for ethnic composition of schools as well as a preference for nearby schools, following Stoica and Flache (2014).

Our main result is that even in the presence of a large proportion of highly tolerant parents, school segregation is likely to arise. Only when all parents have a rather strong preference for nearby schools, do tolerant parents significantly reduce school segregation, through the emergence of mixed schools, a results which is consistent with empirical research (Böhlmark, Holmlund, and Lindahl 2015). Yet, we identified that even under these conditions, the mixed schools were fragile and could tip in the longer run. These phenomena depend on the residential pattern and the location of schools. In particular, lower levels of residential segregation are more favorable to the emergence of mixed schools. Moreover, we found that "complex" residential segregation could make mixed schools more robust to random fluctuations. This occurs because these mixed schools are already located in ethnically diverse areas populated by tolerant parents who find these schools attractive.

We conducted robustness checks to assess whether these conclusions crucially depended on some of our modelling assumptions. We reduced the role of randomness in parental choices (appendix D, figure D.1), increased the number of schools (appendix D, figure D.2), and increased the in-group preference of intolerant parents (appendix D, figure D.3). None of these changes led to qualitatively different results. However, one of our modelling assumptions revealed to be of particular importance: the slight in-group preference of tolerant parents. When making tolerant parents "perfectly tolerant", we found that their desegregating school choices perfectly compensated the actions of intolerant parents even in highly segregated maps (appendix D, figure D.3).

A number of questions remain open for future research. We do not know how robust our findings are when more realistic population settings are considered, especially with one ethnic majority group and several ethnic minority groups with smaller proportions of the population. It seems that ethnic preferences of smaller ethnic minorities are harder to satisfy by integrated schools if those preference resemble those assumed in our model and suggested by empirical studies of residential ethnic preferences (Clark and Fossett 2008).

Another simplification we adopted in this paper was that variation in ethnic preferences was the same in all ethnic groups, with 50% tolerant parents in both groups. But one should also consider the assumption that the distribution of ethnic tolerance varies across ethnic groups. Indeed, empirical research found that natives have different preferences and/or information than immigrants (Kristen 2003), even though a preference for the in-group seems to be common across all groups. This might become even more complex if one would relax the assumption that preferences for nearby schools are similar in all ethnic groups. Indeed, it seems that depending on their socio-economic background parents are unequally able and/or willing to send their children to more distant schools (Andersson, Malmberg, and Östh 2012; Butler et al. 2007). In other words, ethnicity could correlate with tolerance levels and with preference for nearby schools. If mainly parents of one ethnic group prefer mixed schools, the degree to which such schools can remain ethnically mixed might be strongly constrained, in particular if the other group is more willing to pick more distant schools.

While we leave these possible extensions of our model to future research, we believe they all point to changes which may further limit the extent to which the presence of tolerant parents can curb self-reinforcing school segregation on. This suggests that our main finding might turn out to be quite robust, even if a considerable proportion of the population of parents hold very tolerant ethnic preferences there still is a strong potential for unintended preference dynamics to generate substantial levels of school segregation.

Because the presence of a large proportion of tolerant parents alone is not enough to robustly desegregate schools, our analyses suggest that policy makers wishing to reduce school segregation should consider lowering the barriers for desegregating school choices by tolerant parents. Instead of equally targeting the desegregation of all schools, our model suggest to facilitate the extension of the capacity of schools in ethnically diverse neighborhoods while reducing the size of schools in ethnically homogenous neighborhoods. This would allow more tolerant parents to find a school fitting their preferences. However, one should be careful not to make too many intolerant parents unsatisfied, whose actions could then destabilize mixed schools.

It is worth noting that our model mainly applies to situations where school choice is not geographically constrained, like in the Netherlands. Yet, even in countries like France where children's school is assigned on the basis of the residential location, there exist derogations and alternative private school systems where choice is less constrained. Our model suggests that the less constrained are the parental choices, the higher the school segregation, a finding consistent with empirical evidence (Böhlmark et al. 2015). Another policy lever could therefore be to constrain as much as possible parents to send children kids to the closest school from home. However, one should be cautious with this interpretation on the sole basis of our model. Our model only captures "post-residential" school choices (Hastings et al. 2006), assuming that the choice of residential locations is not affected by parents' expectations about the schools available in the proximity of a location. However, empirical studies found that especially ethnic majority parents consider schools when choosing their neighborhood (Bayoh, Irwin, and Haab 2006; Frankenberg 2009), and tend to leave when they are too dissatisfied – the so-called "white flight" phenomenon (Fairlie and Resch 2002; Rangvid 2007; Renzulli and Evans 2005). Therefore, a school choice system that imposes too much constraints could unintendedly foster residential segregation and at the same time fail to reduce school segregation, resonating debates about the unintended consequences of school desegregation policies in the U.S. in the 1960s and 1970s (Logan et al, 2008). A subtler and more locally targeted policies might help to anticipate when schools tend to tip and targeted interventions are needed.

Our findings tentatively suggest that even in a world with increasing proportions of ethnically tolerant parents it remains a formidable challenge to design school choice policies that preclude levels of school segregation most parents neither intend nor desire to bring about. We believe that computational models of the complexities of school choice dynamics have an important part to play in generating the insights needed to master this challenge.

# References

Alba, Richard, and Steven Romalewski. 2013. "The End of Segregation? Hardly.A More Nuanced View from the New York Metropolitan Region." *Center for Urban Research. Www.Urbanresearch.Org*.

Allport, Gordon Willard. 1954. "The Nature of Prejudice." *Addison-Wesley Reading, MA.*

Andersson, Eva, Bo Malmberg, and John Östh. 2012. "Travel-to-School Distances in Sweden 2000–2006: Changing School Geography with Equality Implications." *Journal of Transport Geography* 23:35–43.

Andersson, Roger, Ingar Brattbakk, and Mari Vaattovaara. 2017. "Natives' Opinions on Ethnic Residential Segregation and Neighbourhood Diversity in Helsinki, Oslo and Stockholm." *Housing Studies* 32(4):491–516.

Ashenfelter, Orley, William J. Collins, and Albert Yoon. 2006. "Evaluating the Role of Brown v. Board of Education in School Equalization, Desegregation, and the Income of African Americans." *American Law and Economics Review* 8(2):213–48.

Bakker, J., Denessen, E., Peeters, D., and Walraven, G. (eds.) 2011. *International Perspectives on countering school segregation*. Antwerpen-Apeldoorn, Uitgeverij Garant, ISBN 978 90 4412 6945.

Bayoh, Isaac, Elena G. Irwin, and Timothy Haab. 2006. "Determinants of Residential Location Choice: How Important Are Local Public Goods in Attracting Homeowners to Central City Locations?" *Journal of Regional Science* 46(1):97–120.

Billingham, Chase M., and Matthew O. Hunt. 2016. "School Racial Composition and Parental Choice: New Evidence on the Preferences of White Parents in the United States." *Sociology of Education* 89(2):99–117.

Böhlmark, Anders, Helena Holmlund, and Mikael Lindahl. 2015. *School Choice and Segregation: Evidence from Sweden*. Working Paper.

Borghans, Lex, Bart H. H. Golsteyn, and Ulf Zölitz. 2015. "Parental Preferences for Primary School Characteristics." *The B.E. Journal of Economic Analysis & Policy* 15(1).

Boterman, W., Musterd, S., Pacchi, C., & Ranci, C. (2019). School segregation in contemporary cities: Socio-spatial dynamics, institutional context and urban outcomes. *Urban Studies* 56(15): 3055–3073.

Bruch, Elizabeth E., and Robert D. Mare. 2006. "Neighborhood Choice and Neighborhood Change." *American Journal of Sociology* 112(3):667–709.

Burgess, Simon, Brendon McConnell, Carol Propper, and Deborah Wilson. 2007. "The Impact of School Choice on Sorting by Ability and Socioeconomic Factors in English Secondary Education." *Schools and the Equal Opportunity Problem* 273.

Burgess, Simon, Deborah Wilson, and Ruth Lupton. 2005. "Parallel Lives? Ethnic Segregation in Schools and Neighbourhoods." *Urban Studies* 42(7):1027–56.

Butler, Tim, Chris Hamnett, Mark Ramsden, and Richard Webber. 2007. "The Best, the Worst and the Average: Secondary School Choice and Education Performance in East London." *Journal of Education Policy* 22(1):7–29.

Caetano, Gregorio, and Vikram Maheshri. 2017. "School Segregation and the Identification of Tipping Behavior." *Journal of Public Economics* 148:115–35.

Christ, O., K. Schmid, S. Lolliot, H. Swart, D. Stolle, N. Tausch, A. Al Ramiah, U. Wagner, S. Vertovec, and M. Hewstone. 2014. "Contextual Effect of Positive Intergroup Contact on Outgroup Prejudice." *Proceedings of the National Academy of Sciences* 111(11):3996–4000.

Clark, W. A. V., and M. Fossett. 2008. "Understanding the Social Context of the Schelling Segregation Model." *Proceedings of the National Academy of Sciences* 105(11):4109–14.

Clark, W.A.V. 2015. Residential segregation: Recent trends. In: Wright JD (editor-in-chief) .*International Encyclopedia of the Social & Behavioral Sciences*. 2nd ed. 20. Oxford: Elsevier, pp.549–554. Clark, William AV, Eva K. Andersson, and Bo Malmberg. 2018. "What Can We Learn about Changing Ethnic Diversity from the Distributions of Mixed-Race Individuals?" *Urban Geography* 39(2):263–81.

Clark, William AV, and Noli Brazil. 2019. "Neighborhood Selections by Young Adults: Evidence from a Panel of US Adolescents." *Journal of Urban Affairs* 41(7):981–98.

Clark, William AV, and Regan Maas. 2009. "The Geography of a Mixed-Race Society." *Growth and Change* 40(4):565–93.

Coenders, M. T. A., M. Lubbers, and P. L. H. Scheepers. 2004. "Weerstand Tegen Scholen Met Allochtone Kinderen: De Etnische Tolerantie van Hoger Opgeleiden Op de Proef Gesteld [Resistance against Schools with Allochthonous Pupils: A Test of the Ethnic Tolerance of the Higher Educated]." *147*.

Ellis, Mark, Steven R. Holloway, Richard Wright, and Margaret East. 2007. "The Effects of Mixed-Race Households on Residential Segregation." *Urban Geography* 28(6):554–77.

Ellis, Mark, Steven R. Holloway, Richard Wright, and Christopher S. Fowler. 2012. "Agents of Change: Mixed-Race Households and the Dynamics of Neighborhood Segregation in the United States." *Annals of the Association of American Geographers* 102(3):549–70.

Fairlie, Robert W., and Alexandra M. Resch. 2002. "Is There 'White Flight' into Private Schools? Evidence from the National Educational Longitudinal Survey." *The Review of Economics and Statistics* 84(1):21–33.

Farley, Reynolds, Howard Schuman, Suzanne Bianchi, Diane Colasanto, and Shirley Hatchett. 1978. "'Chocolate City, Vanilla Suburbs:' Will the Trend toward Racially Separate Communities Continue?" *Social Science Research* 7(4):319–44.

Farley, Reynolds, and Alma F. Taeuber. 1974. "Racial Segregation in the Public Schools." *American Journal of Sociology* 79(4):888–905.

Farley, Reynolds, and Karl E. Taeuber. 1968. "Population Trends and Residential Segregation since 1960." *Science* 159(3818):953–56.

Farrell, Chad R., and Barrett A. Lee. 2011. "Racial Diversity and Change in Metropolitan Neighborhoods." *Social Science Research* 40(4):1108–23.

Flache, Andreas, and Rainer Hegselmann. 2001. "Do Irregular Grids Make a Difference? Relaxing the Spatial Regularity Assumption in Cellular Models of Social Dynamics." *Journal of Artificial Societies* *and Social Simulation* 4(4).

Fossett, Mark. 2006. "Ethnic Preferences, Social Distance Dynamics, and Residential Segregation: Theoretical Explorations Using Simulation Analysis." *The Journal of Mathematical Sociology* 30(3–4):185–273.

Frankenberg, Erica. 2009. "The Impact of School Segregation on Residential Housing Patterns: Mobile, Alabama, and Charlotte, North Carolina." Pp. 164–84 in *School resegregation: Must the South turn back?* University of North Carolina Press.

Frankenberg, Erica. 2013. "The Role of Residential Segregation in Contemporary School Segregation." *Education and Urban Society* 45(5):548–70.

Frey, W. 2018. *The Millennial Generation: A Demographic Bridge to America's Diverse Future. Brookings*. Retrieved.

Frey, William H. 2018. *Diversity Explosion: How New Racial Demographics Are Remaking America*. Brookings Institution Press.

Glaeser, Edward, and Jacob Vigdor. 2012. "The End of the Segregated Century: Racial Separation in America's Neighborhoods, 1890–2010." *Manhattan Institute* 36.

Goyette, Kimberly A., Danielle Farrie, and Joshua Freely. 2012. "This School's Gone Downhill: Racial Change and Perceived School Quality among Whites." *Social Problems* 59(2):155–76.

Hall, Matthew, Laura Tach, and Barrett A. Lee. 2016. "Trajectories of Ethnoracial Diversity in American Communities, 1980–2010." *Population and Development Review* 42(2):271–97.

Hastings, Justine S., Thomas J. Kane, and Douglas O. Staiger. 2006. *Preferences and Heterogeneous Treatment Effects in a Public School Choice Lottery. Working Paper*. 12145. National Bureau of Economic Research.

Hatna, Erez, and Itzhak Benenson. 2015. "Combining Segregation and Integration: Schelling Model Dynamics for Heterogeneous Population." *Journal of Artificial Societies and Social Simulation* 18(4):15.

Hegselmann, Rainer. 2017. "Thomas C. Schelling and James M. Sakoda: The Intellectual, Technical, and Social History of a Model." *Journal of Artificial Societies and Social Simulation* 20(3):15.

Ibraimovic, Tatjana, and Lorenzo Masiero. 2014. "Do Birds of a Feather Flock Together? The Impact of Ethnic Segregation Preferences on Neighbourhood Choice." *Urban Studies* 51(4):693–711.

Johnson, Rucker C. 2011. *Long-Run Impacts of School Desegregation & School Quality on Adult Attainments*. National Bureau of Economic Research.

Johnston, Ron, Simon Burgess, Deborah Wilson, and Richard Harris. 2006. "School and Residential Ethnic Segregation: An Analysis of Variations across England's Local Education Authorities." *Regional Studies* 40(9):973–90.

Karsten, Sjoerd, Guuske Ledoux, Jaap Roeleveld, Charles Felix, and DorothÉ Elshof. 2003. "School Choice and Ethnic Segregation." *Educational Policy* 17(4):452–77.

Kessel, Dany, and Elisabet Olme. 2018. "School Choice Priority Structures and School Segregation."

*Unpublished Manuscript* 44.

Kim, Jae Hong, Francesca Pagliara, and John Preston. 2005. "The Intention to Move and Residential Location Choice Behaviour." *Urban Studies* 42(9):1621–36.

Kristen, Cornelia. 2003. *School Choice and Ethnic School Segregation: Primary School Selection in Germany*. Waxmann Verlag.

Lee, Barrett A., John Iceland, and Gregory Sharp. 2012. "Racial and Ethnic Diversity Goes Local: Charting Change in American Communities Over Three Decades | RSF." Retrieved April 8, 2020 (https://www.russellsage.org/research/reports/racial-ethnic-diversity).

Massey, Douglas S., and Nancy A. Denton. 1988. "The Dimensions of Residential Segregation." *Social Forces* 67(2):281.

McFadden, Daniel. 1973. "Conditional Logit Analysis of Qualitative Choice Behavior."

Okabe, Atsuyuki, B. N. Boots, and Kōkichi Sugihara. 1992. *Spatial Tessellations: Concepts and Applications of Voronoi Diagrams*. Chichester, England ; New York: John Wiley & Sons Ltd.

Ong, Paul M., and Jordan Rickles. 2004. "The Continued Nexus between School and Residential Segregation." *Asian LJ* 11:260.

Paolillo, Rocco, and Jan Lorenz. 2018. "How Different Homophily Preferences Mitigate and Spur Ethnic and Value Segregation: Schelling's Model Extended." *Advances in Complex Systems* 21(06n07):1850026.

Powers, Daniel A., and Christopher G. Ellison. 1995. "Interracial Contact and Black Racial Attitudes: The Contact Hypothesis and Selectivity Bias." *Social Forces* 74(1):205–26.

Rangvid, Beatrice Schindler. 2007. "Living and Learning Separately? Ethnic Segregation of School Children in Copenhagen." *Urban Studies* 44(7):1329–54.

Reardon, Sean F., and Ann Owens. 2014. "60 Years after Brown: Trends and Consequences of School Segregation." *Annual Review of Sociology* 40:199–218.

Reardon, Sean F., and John T. Yun. 2005. "Integrating Neighborhoods, Segregating Schools: The Retreat from School Desegregation in the South, 1990-2000." *School Resegregation: Must the South Turn Back?*

Renzulli, Linda A., and Lorraine Evans. 2005. "School Choice, Charter Schools, and White Flight." *Social Problems* 52(3):398–418.

van de Rijt, Arnout, David Siegel, and Michael Macy. 2009. "Neighborhood Chance and Neighborhood Change: A Comment on Bruch and Mare." *American Journal of Sociology* 114(4):1166–80.

Sakoda, James M. 1971. "The Checkerboard Model of Social Interaction." *The Journal of Mathematical Sociology* 1(1):119–32.

Schelling, Thomas C. 1971. "Dynamic Models of Segregation." *The Journal of Mathematical Sociology* 1(2):143–86.

Schwanen, Tim, and Patricia L. Mokhtarian. 2004. "The Extent and Determinants of Dissonance

between Actual and Preferred Residential Neighborhood Type." *Environment and Planning B: Planning and Design* 31(5):759–84.

Skvoretz, John. 2006. "Introduction to Special Edition on 'Ethnic Preferences, Social Distance Dynamics, and Residential Segregation: Theoretical Explanations Using Simulation Analysis' by Mark Fossett." *The Journal of Mathematical Sociology* 30(3–4):181–84.

Spaiser, Viktoria, Peter Hedström, Shyam Ranganathan, Kim Jansson, Monica K. Nordvik, and David J. T. Sumpter. 2018. "Identifying Complex Dynamics in Social Systems: A New Methodological Approach Applied to Study School Segregation." *Sociological Methods & Research* 47(2):103–35.

Stoica, Victor Ionut, and Andreas Flache. 2014. "From Schelling to Schools: A Comparison of a Model of Residential Segregation with a Model of School Segregation." *Journal of Artificial Societies and Social Simulation* 17(1):5.

Tasan-Kok, Tuna, Ronald Van Kempen, Raco Mike, and Gideon Bolt. 2014. *Towards Hyper-Diversified European Cities: A Critical Literature Review*. Utrecht University.

Vertovec, Steven. 2007. "Super-Diversity and Its Implications." *Ethnic and Racial Studies* 30(6):1024–54.

Wölfer, Ralf, Katharina Schmid, Miles Hewstone, and Maarten van Zalk. 2016. "Developmental Dynamics of Intergroup Contact and Intergroup Attitudes: Long-Term Effects in Adolescence and Early Adulthood." *Child Development* 87(5):1466–78.

Xie, Y., and X. Zhou. 2012. "Modeling Individual-Level Heterogeneity in Racial Residential Segregation." *Proceedings of the National Academy of Sciences* 109(29):11646–51.

Zhang, Junfu. 2004. "Residential Segregation in an All-Integrationist World." *Journal of Economic Behavior & Organization* 54(4):533–50.

**Appendix A: Residential model**

To generate the residential maps over which the school choice model is launched, we use two Schelling-Sakoda type of algorithms, in which agents relocate themselves according to their ethnic preferences only. In other words, distance plays no role, or more formally in equation (1): $\alpha = 1$. We thus have:

$$p_{ijt+1} = \frac{e^{\beta S_g(x_{jt})}}{\sum_{k=1}^{K} e^{\beta S_g(x_{kt})}} \tag{6}$$

With:

$$\sum_{k=1}^{K} p_{ijt+1} = 1 \tag{7}$$

where $p_{ijt+1}$ is the probability for agent $i$ of ethnic group $g$, to move to cell $j$ at time $t+1$ (the next round of the simulation); $\beta$ is the same parameter as in the school choice model that manipulates the weight of the ethnic satisfaction $S$ relative to random chance; $S_g(x_{jt})$ is given by equation (2), with $x_{jt}$ representing the proportion of in-group parents in the neighborhood around the empty cell. Instead of evaluating the 30 schools, agents evaluate 8 available locations picked at random in all existing empty spots (the map contains 10% of empty cells), to which they add their current location (K = 9 cells in total). With this procedure, the more attractive potential housing locations are the more likely to be chosen.

For each cell evaluated agents compute the ethnic ratio of the neighborhood composed by all cells in a diamond-shape with radius equal to 6 (112 cells).

One advantage of this modeling strategy is that it is better able to model diversity in (ethnic) preferences than with thresholds. Even though it is possible to model diversity in tolerance with threshold functions by varying threshold levels in the population (Hatna and Benenson 2015; Paolillo and Lorenz 2018), this results in more tolerant agents moving less than others. Without noise, that is, agents moving at random with a certain probability at each tick, the cascading dynamic of movements typical of Schelling residential segregation's models would never be triggered, because "tolerant" agents would not move, and the model would get stuck in its initial conditions. Indeed, with thresholds, whatever the functional form of the utility function, everything that is above the threshold is equally satisfying and everything below it is equally dissatisfying. Thus, the decisions of agents do not smoothly reflect their preferences: the agents do not look for better situations than what they already have if they are "satisfied" already. Since we wish to model tolerant preferences, not only as a lower need for similar people, but also as a light aversion for too much similarity, we use the combination of single-peaked utility functions (Zhang 2004), and random-utility model.

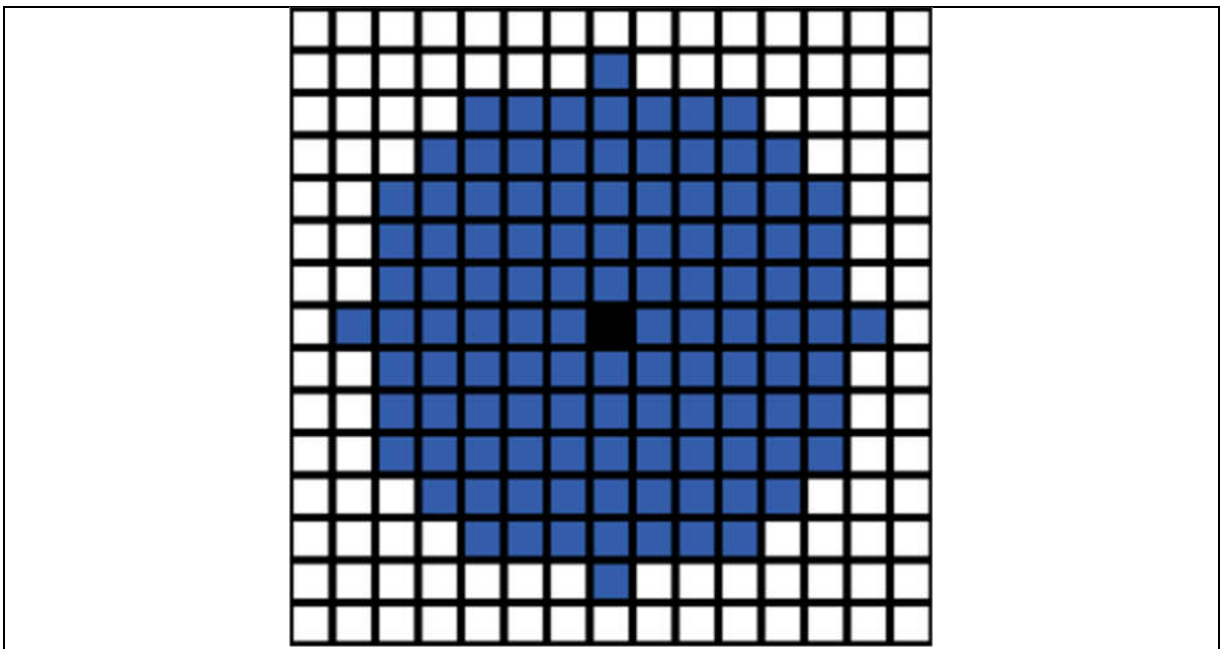

<u>Figure A:</u> Neighborhood of radius 6 and diamond shape. When the empty cell (in black in the center) is selected by an agent as one of the 8 potential destinations (plus his current one), he computes the ratio of agents of the same ethnicity as him on the total number of agents in the blue cells.

The two map generator methods only differ with respect to the functional form given to the ethnic satisfaction function $S$. In map generator method 2, we use the same parametrization as in the school choice model, while in map generator method 1 all parents have the ethnic satisfaction function $S$ of intolerant parents $x_0 = 0.8$ and $M = 0.6$. In this case, half of the parents "become" tolerant in the school choice model only, that is, once the residential locations has been chosen and is fixed.

**Appendix B: The tipping of schools in the long run**

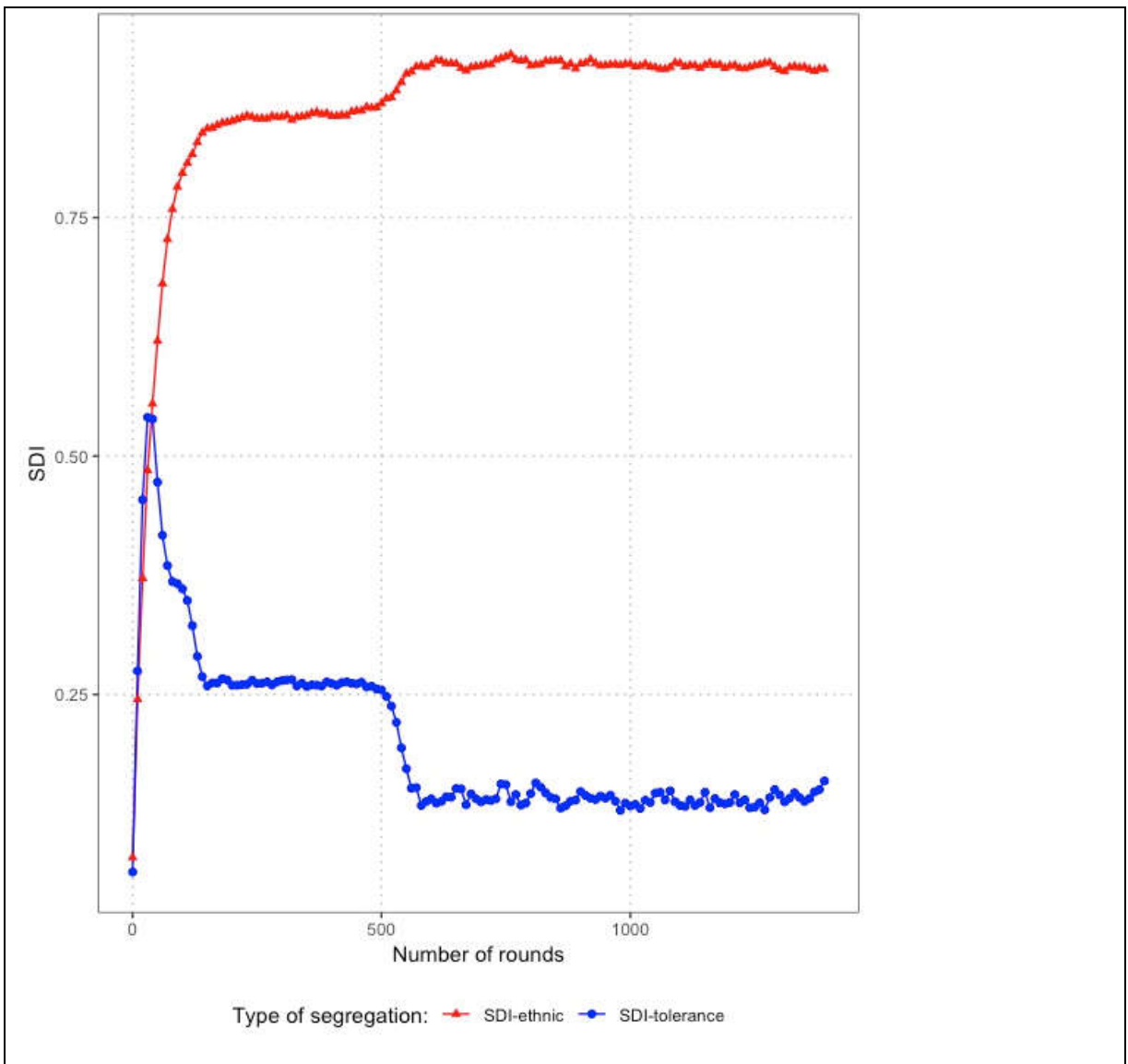

<u>Figure B:</u> Evolution of school segregation indices over one simulation run of 1400 rounds starting from integrated map ($RDI \leq 0.25$); $\alpha = 1$.

This figure illustrates the instability of integrated schools without distance constraint (same parametrization as in figure 6 but in the longer run). At around 500 rounds we clearly see that an integrated school tips, which instantaneously reduces segregation by tolerance and increases ethnic segregation even though the run had been stable for more than 300 rounds.

**Appendix C: Comparison between ethnic school segregation between "simple" and "complex" maps**

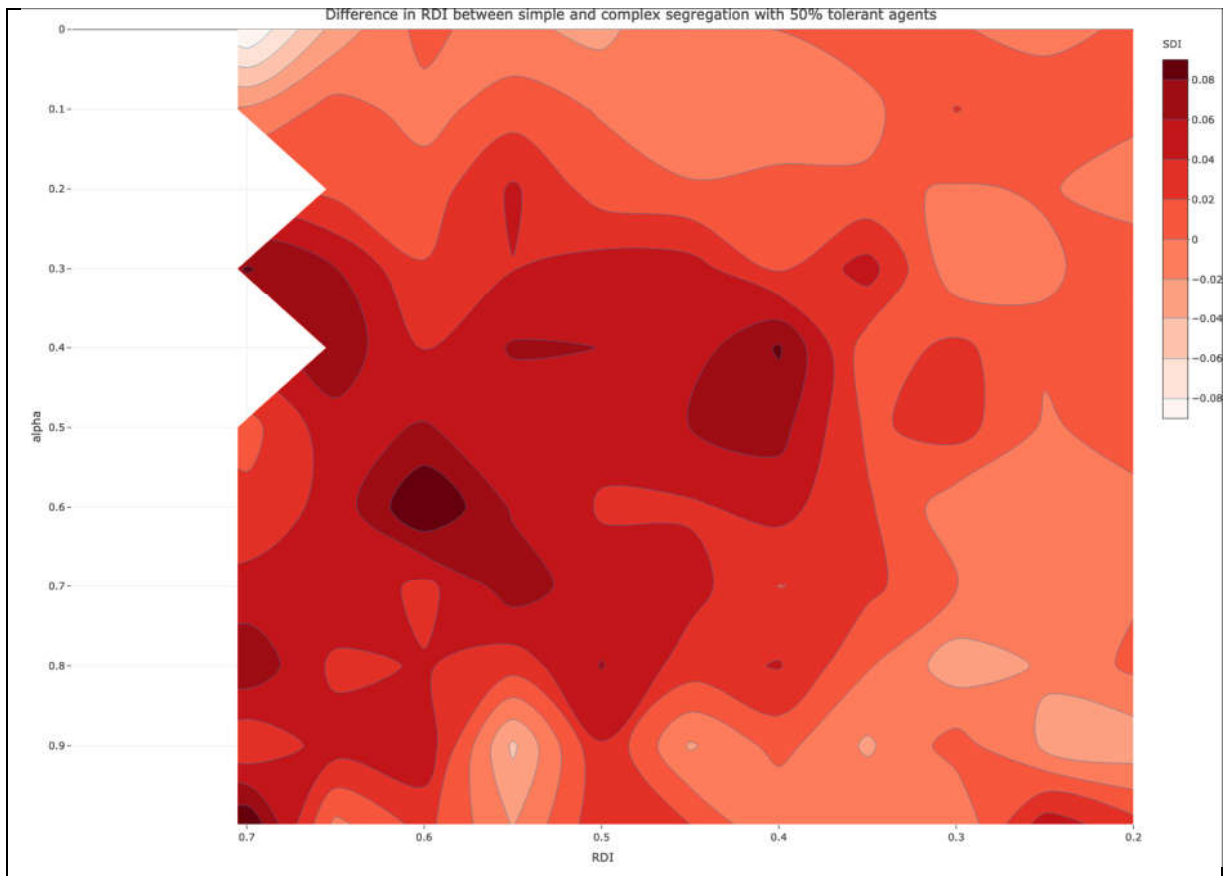

Figure C: Difference in school dissimilarity index SDI (z-axis), between "simple" and "complex" forms of segregation, depending on residential dissimilarity index RDI (x-axis), $\alpha$ (y-axis), with 50% tolerant agents. The surface is drawn based on 4686 observations each corresponding to one complete simulation run outcome. Since RDI is continuous we split it into intervals of width 0.05. We average all values of SDI that correspond to each combination of $\alpha$ and RDI separately for "simple" and "complex" continuums. We then compute the difference between the values obtained in each case.
https://plot.ly/~lsage/29/#/

The graph is hard to read. Two major things can be noticed: first, as expected, there a no clear difference between the effect of tolerant agents on school segregation for low levels of residential segregation. Second, the "complex" form of segregation seem to lead to lower levels of school segregation for high levels of residential segregation, even though this effect is moderate.

**Appendix D: Robustness checks with "simple" segregation**

In the results presented in this appendix we varied some of parameters held constant in the main analyses. Since for $\alpha = 0$ (parents only consider the distance between home and school) there is no change across experimental conditions, in the robustness checks presented in this appendix we did not run new simulations for these points of the parameter space, and instead used the same values as in the main analyses of the article. All figures of this appendix should be compared to figure 3 in the main analyses.

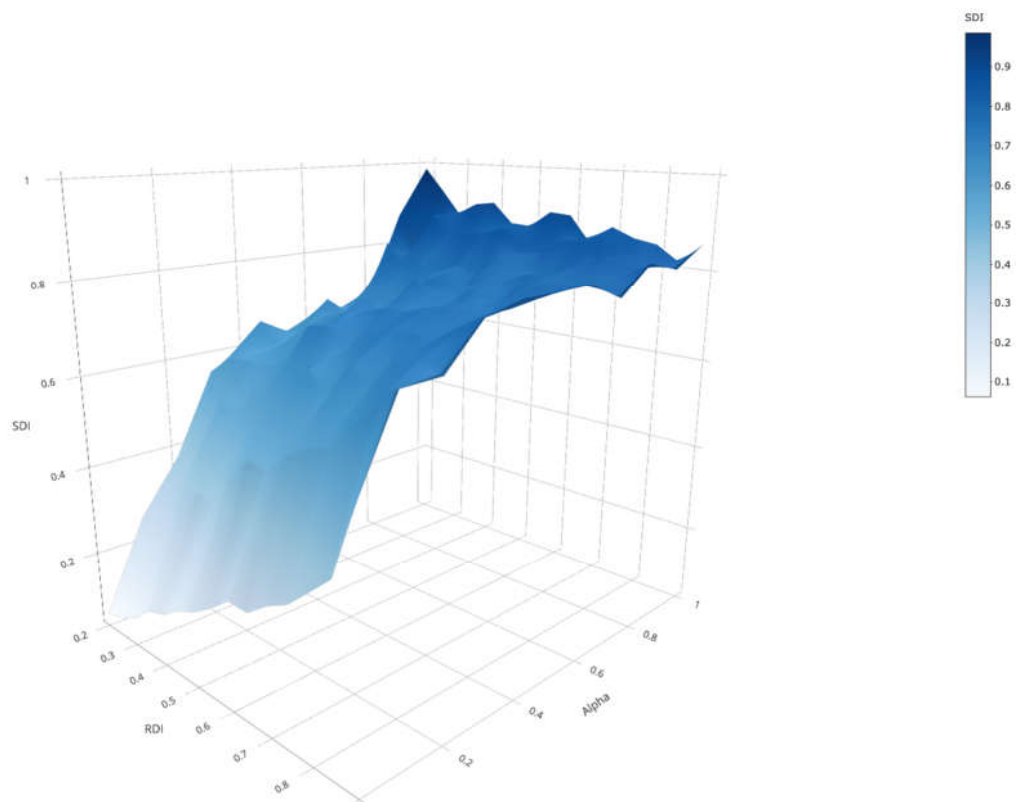

<u>Figure D.1</u>: $\beta = 100$. Relations between residential dissimilarity index RDI (x-axis), $\alpha$ (y-axis) and school dissimilarity index SDI (z-axis), for 50% tolerant agents, for the "segregated" maps continuum generated with map generator method 1. The surface is drawn based on 2100 original observations each corresponding to one complete simulation run outcome. Since RDI is continuous we split it into intervals of width 0.05. We average all values of SDI that correspond to each combination of $\alpha$ and RDI.
https://plot.ly/~lsage/78/#/

The higher $\beta$, which leaves less room for randomness in parent's choices, does not qualitatively change the relationships between our variables of interest.

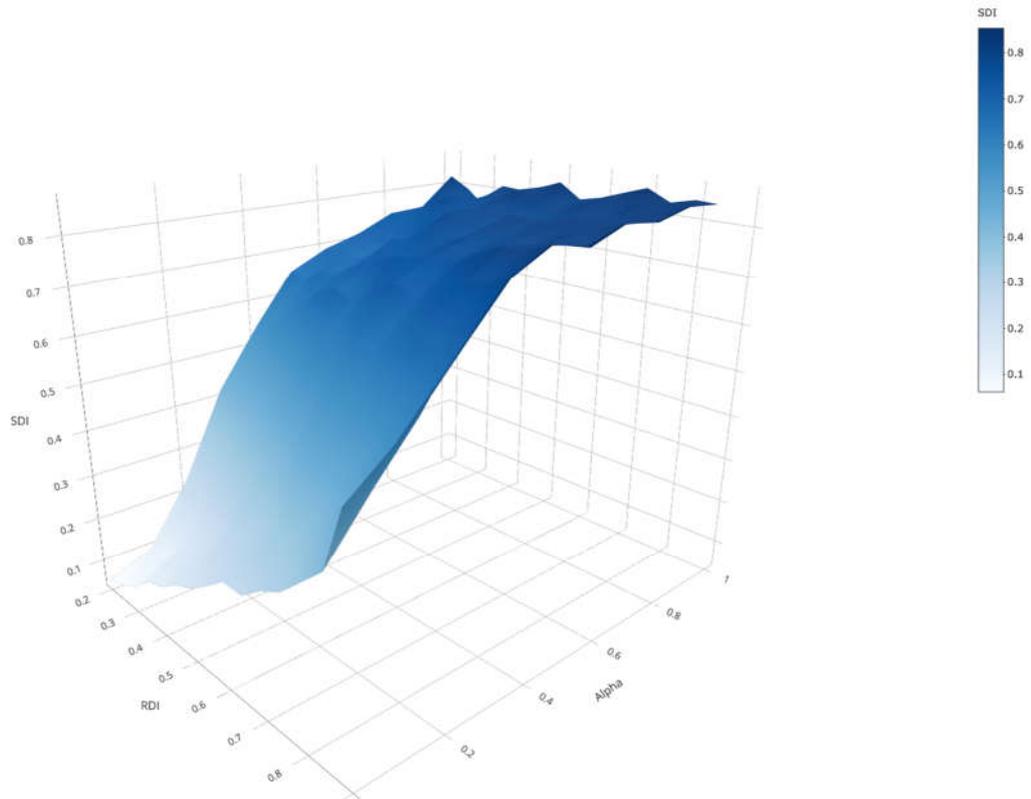

Figure D.2: Number of schools = 40. Relations between residential dissimilarity index RDI (x-axis), α (y-axis) and school dissimilarity index SDI (z-axis), for 50% tolerant agents, for the "segregated" maps continuum generated with map generator method 1. The surface is drawn based on 2100 original observations each corresponding to one complete simulation run outcome. Since RDI is continuous we split it into intervals of width 0.05. We average all values of SDI that correspond to each combination of α and RDI.
https://plot.ly/~lsage/72/#/

Adding 10 more schools does not seem to affect our results in the short run.

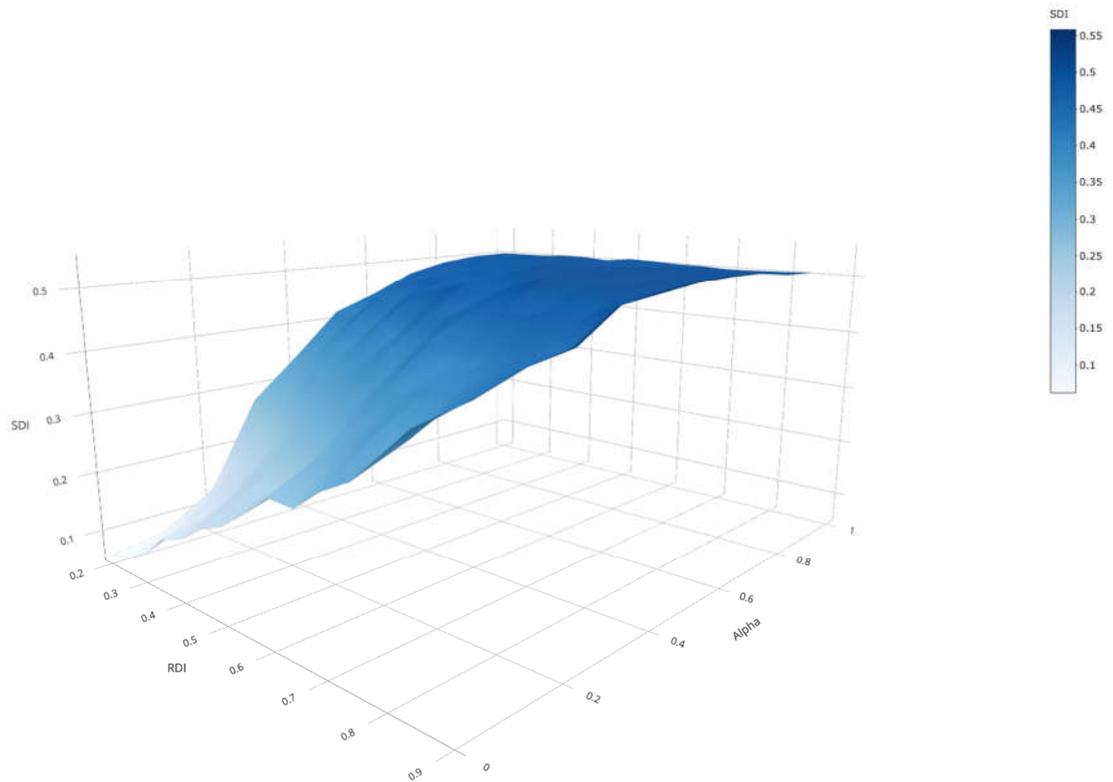

<u>Figure D.3</u>: $M\ tolerant = 0$. Relations between residential dissimilarity index RDI (x-axis), $\alpha$ (y-axis) and school dissimilarity index SDI (z-axis), for 50% tolerant agents, for the "segregated" maps continuum generated with map generator method 1. The surface is drawn based on 2100 original observations each corresponding to one complete simulation run outcome. Since RDI is continuous we split it into intervals of width 0.05. We average all values of SDI that correspond to each combination of $\alpha$ and RDI.
https://plot.ly/~lsage/74/#/

When parents are perfectly tolerant, we see that they perfectly compensate the segregating force of intolerant parents. We see no conditions under which ethnic segregation exceeds $SDI = 0.5$.

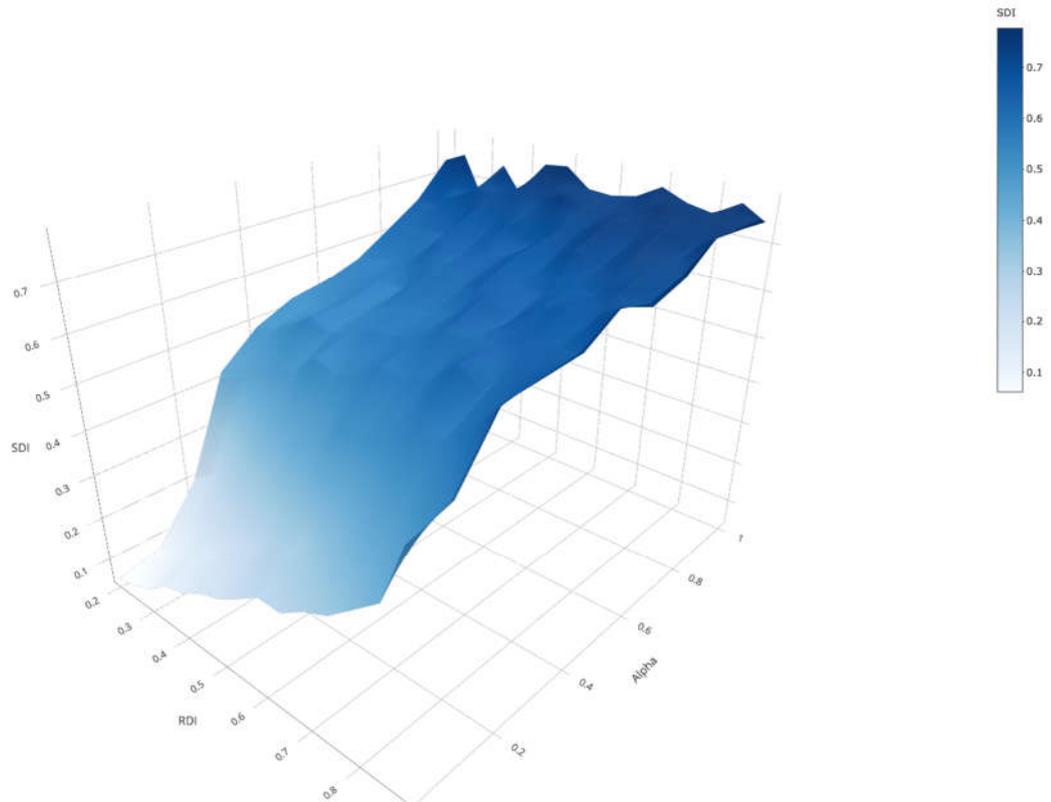

<u>Figure D.4</u>: $x_0\ intolerant = 1$. Relations between residential dissimilarity index RDI (x-axis), $\alpha$ (y-axis) and school dissimilarity index SDI (z-axis), for 50% tolerant agents, for the "segregated" maps continuum generated with map generator method 1. The surface is drawn based on 2100 original observations each corresponding to one complete simulation run outcome. Since RDI is continuous we split it into intervals of width 0.05. We average all values of SDI that correspond to each combination of $\alpha$ and RDI.
https://plot.ly/~lsage/76/#/

When intolerant parents are maximally intolerant $x_0 = 1$, our results are not affected in the short run.